\documentclass[a4paper,11pt]{article}

\usepackage{jcappub} 

\usepackage{graphicx}  
\usepackage{dcolumn}
\usepackage{tabularx}
\usepackage{bm}        
\usepackage{amssymb,amsfonts,amsmath,physics}   
\usepackage[dvipsnames]{xcolor}
\usepackage{slashed}
\usepackage{multirow}
\usepackage{dsfont}

\usepackage{orcidlink}
\usepackage{comment}
\usepackage{array}
\usepackage[normalem]{ulem}
\newcolumntype{L}[1]{>{\raggedright\let\newline\\\arraybackslash\hspace{0pt}}m{#1}}
\newcolumntype{C}[1]{>{\centering\let\newline\\\arraybackslash\hspace{0pt}}m{#1}}
\newcolumntype{R}[1]{>{\raggedleft\let\newline\\\arraybackslash\hspace{0pt}}m{#1}}

\def\bra {\langle}
\def\ket {\rangle}
\def\to {\rightarrow}

\def\r {\rightarrow}

\def\bar {\overline}

\def\lr {\leftrightarrow}

\def\bra {\langle}
\def\ket {\rangle}
\def\abs#1{\left| #1 \right|}
\def\braket#1{\bra #1 \ket}

\newcommand{\neff}{N_{\rm eff}}

\newcommand{\bmt}{\begin{pmatrix}}
\newcommand{\emt}{\end{pmatrix}}
\newcommand{\ba}{\begin{array}{c}}
\newcommand{\ea}{\end{array}}
\newcommand{\be}{\begin{equation}}
\newcommand{\ee}{\end{equation}}
\newcommand{\bea}{\begin{eqnarray}}
\newcommand{\eea}{\end{eqnarray}}

\newcommand{\bi}{\begin{itemize}}
\newcommand{\ei}{\end{itemize}}

\newcommand{\baz}{\begin{array}{cc}}
\newcommand{\mathsym}[1]{{}}

\newcommand{\bt}{\begin{tabular}}
\newcommand{\et}{\end{tabular}}

\newcommand{\benu}{\begin{enumerate}}
\newcommand{\eenu}{\end{enumerate}}

\title{\boldmath  
	 Large neutrino mass in cosmology and keV sterile neutrino dark matter from a dark sector}
  
\author[a,b]{Cristina Benso\orcidlink{0000-0003-1922-8534},}
\author[b]{Thomas Schwetz\orcidlink{0000-0001-7091-1764}}
\author[c,d]{and Drona Vatsyayan\orcidlink{0000-0002-6868-3237}}

\affiliation[a]{Institute for Theoretical Particle Physics (TTP), Karlsruhe Institute of Technology (KIT), 76128 Karlsruhe, Germany}

\affiliation[b]{Institut für Astroteilchen Physik, Karlsruher Institut für Technologie (KIT), Hermann-von-Helmholtz-Platz 1,
76344 Eggenstein-Leopoldshafen, Germany}

\affiliation[c]{Departament de Física Teòrica, Universitat de València, 46100 Burjassot, Spain}

\affiliation[d]{Instituto de Física Corpuscular (CSIC-Universitat de València),
Parc Científic UV,\\ C/Catedrático José Beltrán, 2, E-46980 Paterna, Spain}

\emailAdd{cristina.benso@kit.edu}
\emailAdd{schwetz@kit.edu}
\emailAdd{drona.vatsyayan@ific.uv.es}

\abstract{We consider an extended seesaw model which generates active neutrino masses via the usual type-I seesaw and leads to a large number of massless fermions as well as a sterile neutrino dark matter (DM) candidate in the $\mathcal{O}(10-100) {\rm~keV}$ mass range. The dark sector comes into thermal equilibrium with Standard Model neutrinos after neutrino decoupling and before recombination via a U(1) gauge interaction in the dark sector. This suppresses the abundance of active neutrinos and therefore reconciles sizeable neutrino masses with cosmology. The DM abundance is determined by freeze-out in the dark sector, which allows avoiding bounds from X-ray searches. Our scenario predicts a slight increase in the effective number of neutrino species $N_{\rm eff}$ at recombination, potentially detectable by future CMB missions.}


\arxivnumber{2410.23926}

\begin{document} 
\maketitle
\flushbottom

\section{Introduction}
\label{sec:intro}

Non-zero neutrino masses leave imprints on cosmological observables, which in turn can be used to constrain their properties in several cosmological surveys \cite{Lesgourgues:2006nd}. In combination with \emph{Planck} cosmic microwave observations~\cite{Planck:2018vyg}, the latest results from the DESI collaboration \cite{DESI:2024mwx} place a stringent bound on the sum of neutrino masses in the standard $\Lambda$CDM model:  
\begin{equation}\label{eq:cosmobound}
    \sum m_\nu \equiv \sum_{i=1}^3 m_i < 0.072 {\rm~eV}~ (95\%{\rm~CL}),
\end{equation}
which can be compared with the earlier bound from the \emph{Planck} collaboration $\sum m_\nu < 0.12 {\rm~eV}~ (95\%{\rm~CL})$ \cite{Planck:2018vyg}. As these bounds from cosmology are getting stronger and stronger, they are nearing or already in tension with the laboratory constraints on the same. In particular, neutrino oscillation data put a lower limit on the sum of neutrino masses: $\sum m_\nu > 0.058~(0.098) {\rm~eV}$ (95\%~CL) for normal (inverted) neutrino mass ordering, when the lightest neutrino is massless ($m_{\rm lightest} \r 0$)~\cite{Esteban:2024eli}, implying that the DESI bound is in tension with inverted ordering at the $2\sigma$ level. Recently, it has even been argued that cosmology would prefer a negative effective neutrino mass \cite{Craig:2024tky,Green:2024xbb,Elbers:2024sha} (see, however, also \cite{Naredo-Tuero:2024sgf,Loverde:2024nfi,RoyChoudhury:2024wri}), suggesting that these bounds might become even more stringent in the future, leading to significant discord between the cosmology and neutrino oscillations \cite{Gariazzo:2023joe}.

At the same time, the current best model-independent limit on the neutrino mass measurement from the KATRIN experiment gives $m_\beta < 0.45 {\rm~eV}$ \cite{Katrin:2024tvg}, which would correspond to $\sum m_\nu \lesssim 1.35 {\rm~eV}$.  Similarly, the search for neutrinoless double beta decay ($0\nu\beta\beta$) to prove the Majorana nature of neutrinos at the KamLand-Zen experiment also gives a constraint on the effective Majorana mass, $m_{\beta\beta} < 0.028 - 0.122 {\rm~eV}~ (90\% {\rm~CL})$ \cite{KamLAND-Zen:2024eml}, 
which gives $m_{\rm lightest} < 0.084 - 0.353 {\rm~eV}$ at 90\% CL.

Hence, if our universe is well described by the standard cosmological scenario, we might not be able to observe finite absolute neutrino mass in the laboratory. If we are to expect positive results from the ongoing experiments or if cosmological constraints become significantly inconsistent with neutrino oscillations, we need a mechanism to reconcile the bounds from cosmology and laboratory. In the literature, several non-standard scenarios to relax the cosmological mass bound have been proposed. For example, they invoke neutrino decays \cite{Escudero:2019gfk,Chacko:2019nej,Escudero:2020ped,Chacko:2020hmh,Barenboim:2020vrr,FrancoAbellan:2021hdb,Chen:2022idm}, time-varying neutrino masses \cite{Dvali:2016uhn,Lorenz:2018fzb,Dvali:2021uvk,Lorenz:2021alz,Esteban:2021ozz,Sen:2023uga}, presence of dark radiation \cite{Escudero:2022gez,Farzan:2015pca,Allali:2024anb,GAMBITCosmologyWorkgroup:2020htv}, and neutrinos with distribution different from Fermi-Dirac distribution \cite{Cuoco:2005qr,Alvey:2021sji,Oldengott:2019lke}.  Phenomenological implications of some of these so-called \textit{large-neutrino mass cosmologies} have been discussed in ref.~\cite{Alvey:2021xmq}. 

Another cosmological puzzle that can be related to the neutrino sector is the problem of dark matter (DM). Indeed, if sterile neutrinos exist in nature, they can be considered viable dark matter candidates, see refs.~\cite{Drewes:2016upu,Abazajian:2017tcc,Boyarsky:2018tvu} for reviews.  While keV sterile neutrino DM produced via the Dodelson-Widrow mechanism is excluded, there have been attempts in the literature to modify the standard picture to accommodate sterile neutrino DM by introducing active neutrino self-interactions \cite{DeGouvea:2019wpf,Kelly:2020pcy,Benso:2021hhh} and sterile neutrino self-interactions \cite{Johns:2019cwc,Bringmann:2022aim,Astros:2023xhe,Fuller:2024noz}, see also \cite{Datta:2021elq,Goertz:2024gzw}. Moreover, in refs.~\cite{Berlin:2017ftj,Berlin:2018ztp}, it was shown that it is possible to have thermal DM below the MeV scale, if the DM comes into equilibrium with the standard model (SM) below the temperature of neutrino-photon decoupling.
 
In this work, we present a way to relax the cosmological neutrino mass bound with the assistance of a dark sector comprising dark radiation and sterile neutrino dark matter (DM) by employing self-interactions of active and sterile states. We depart from the model from ref.~\cite{Escudero:2022gez}, which is based on a \textit{Minimal Extended Seesaw} scenario~\cite{Barry:2011wb} and realises a mechanism formulated by Farzan and Hannestad \cite{Farzan:2015pca} to allow for large neutrino masses. We show that this model can be extended in a straight forward way to accommodate a keV scale DM candidate, whose cosmic abundance is obtained via a freeze-out mechanism in the dark sector,
reconciling it with constraints from astrophysical and cosmological observations.
 
The outline of the work is as follows: In Sec.~\ref{sec:mech}, we review the mechanism to relax the cosmological neutrino mass bound with a light dark sector. We then present the model to realize the mechanism in Sec.~\ref{sec:model}. The dark matter phenomenology is discussed in Sec.~\ref{sec:dm} and the constraints on the parameter space of the model, suppression of the neutrino mass as well as predictions for the effective relativistic degrees of freedom (i.e., the effective number of neutrino species $N_{\rm eff}$) are provided in Sec.~\ref{sec:neutrino-mass}. Finally, we conclude in Sec.~\ref{sec:conc} and provide some appendices with additional details.

\section{Relaxing the cosmological neutrino mass bound due to a light dark sector}
\label{sec:mech}

We employ the mechanism to relax the cosmological bound on the neutrino mass introduced in \cite{Farzan:2015pca} and further elaborated in \cite{Escudero:2022gez}. Here, we review it briefly and discuss how it is modified in the presence of a DM candidate. 

The bound on the sum of neutrino masses in eq.~\eqref{eq:cosmobound} is derived indirectly from the cosmological observations, which are sensitive to the energy density in  neutrinos. The CMB and large-scale structure observations, which are insensitive to the exact distribution function of neutrinos~\cite{Alvey:2021sji}, place a bound only on the energy density of non-relativistic neutrinos \cite{Planck:2018vyg}
\begin{equation}
    \Omega_\nu h^2 \equiv \frac{\sum m_\nu n_\nu^0 h^2}{\rho_{\rm critical}} < 1.3 \times 10^{-3}~(95\% {\rm~CL})\,.
\end{equation}
As can be seen from the equation above, the cosmological observations thus place a bound on the product of neutrino masses and their number density
\begin{equation}
    \sum m_\nu \times \left(\frac{n_\nu^0}{56~ {\rm cm}^{-3}}\right) < 0.12 {\rm~eV}~ (95\%{\rm~CL})\,,
\end{equation}
with $n_\nu^0$ denoting the background neutrino number density per helicity state. For illustration purposes, here we use the bound from \emph{Planck} \cite{Planck:2018vyg}.
Hence, if we introduce a mechanism that can reduce the number density of neutrinos, larger neutrino masses can be accommodated.  

At earlier times, when neutrinos are ultra-relativistic, their energy density is characterized by the number of effective ultra-relativistic neutrino species,  $N_{\rm eff}$
\begin{equation}
    N_{\rm eff} \equiv \frac{8}{7}\left(\frac{11}{4}\right)^{4/3}\,\left(\frac{\rho_{\rm rad}-\rho_\gamma}{\rho_\gamma} \right)\,,
\end{equation}
where $\rho_{\rm rad}$ is the total energy density in relativistic species, and we have $N_{\rm eff} \propto \braket{p_\nu} n_\nu $. Since the current measurement of $\neff = 2.99 \pm 0.17$ \cite{Planck:2018vyg} is in good agreement with the standard model (SM) calculation $\neff^{\rm SM} = 3.044(1)$ \cite{EscuderoAbenza:2020cmq,Akita:2020szl}, the decrease in $n_\nu$ must be compensated by the addition of light/massless beyond standard model species, and both this reduction of neutrino number density and addition of BSM states should happen before recombination. Moreover, in order to not spoil the successful predictions of Big Bang Nucleosyntesis (BBN), the mechanism should be activated once proton-neutron conversions have frozen-out around $T_\gamma \sim 0.7 {\rm~MeV}$. Therefore, the mechanism would involve BSM states that thermalize with the neutrinos in the period between BBN and recombination, 10~eV $\lesssim T_\gamma \lesssim 100$~keV, so that $\neff$ remains unchanged at earlier times.

As the neutrinos decouple at $T_\gamma \sim 2$~MeV, they cannot be produced anymore from the particles in the SM thermal plasma. This implies that the production of new states would happen at their expense, thus reducing their number density and consequently relaxing the cosmological mass bound, such that cosmology becomes sensitive to the quantity
\begin{equation}\label{eq:sum-suppression}
    \left[\sum m_\nu\right]_{\rm eff} =
    \sum m_\nu \, \frac{n_\nu}{n_\nu^{\rm SM}} \,,
\end{equation}
where $n_\nu^{\rm SM}$ is the neutrino number density as predicted in the standard $\Lambda$CDM model. Hence, for $n_\nu < n_\nu^{\rm SM}$, a large value of $\sum m_\nu$ can be accommodated. The suppression factor depends on the massive and massless degrees of freedom in the dark sector, and we will discuss its values in our model in sec.~\ref{sec:neff} below.

Ref.~\cite{Escudero:2022gez} realized this mechanism by considering a new boson $X$ with mass $100{\rm~eV} \lesssim m_X \lesssim 1 {\rm~MeV} $, that thermalizes with neutrinos and numerous massless BSM species ($\chi$) via processes such as $\nu \bar{\nu} \r X$ and $X \r \chi\chi$, with the requirements $\braket{\Gamma(\nu \bar{\nu} \r X)} > H$ and $\braket{\Gamma(X \r \chi_i + \chi_j {\rm~or~} \nu)} > H$ in the temperature window 10 eV $ \lesssim T \lesssim$ 1 MeV, where $H$ is the Hubble expansion rate. This reduces the neutrino population to produce the new boson, which then decays dominantly into the massless states $\chi$, thereby populating the dark radiation component at the expense of the massive SM neutrinos. \\

\begin{figure*}[!t]
    \centering
    \includegraphics[width=0.98\linewidth]{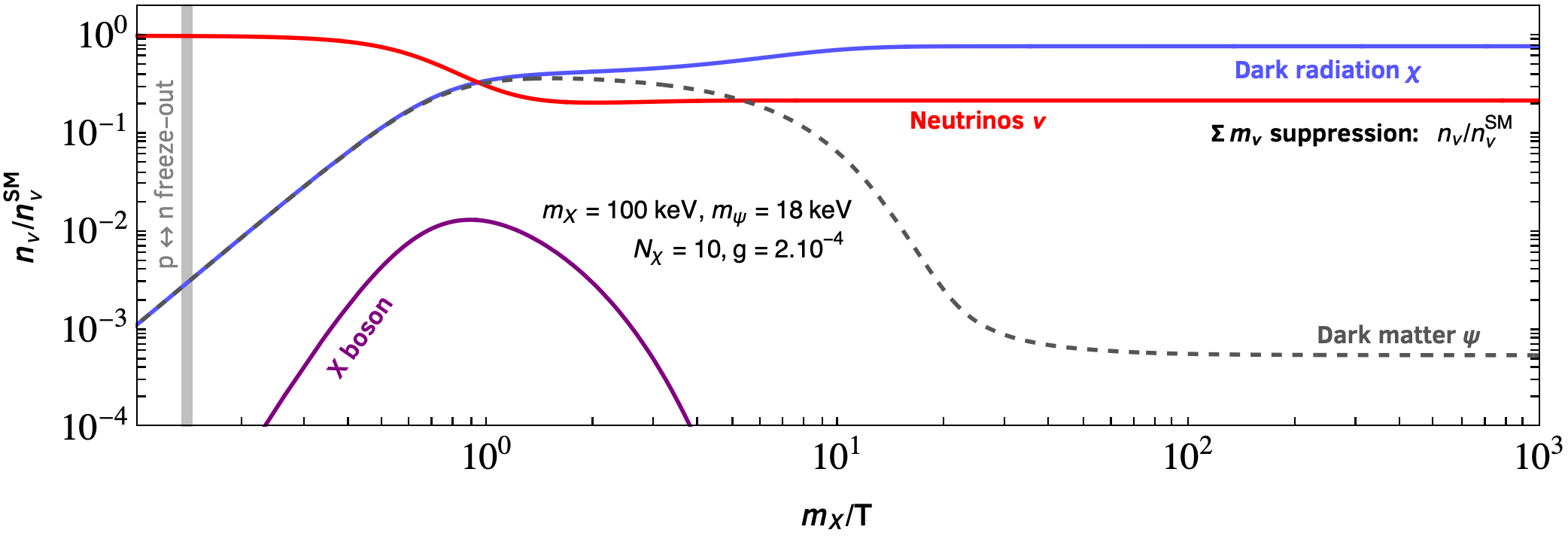}
    \caption{Schematic representation of the mechanism to reduce the active neutrino number density relative to the one in the Standard Model (shown in red) by equilibrating a dark sector consisting of $N_\chi = 10$ massless Dirac fermion species and the DM candidate $\psi$ with mass $m_\psi = 18$~keV due to a mediator $X$ with mass $m_X = 100$~keV and coupling $g = 2 \times 10^{-4}$. The $X$ boson is produced initially from the inverse decays $\nu \bar{\nu} \r X$ and then decays dominantly to dark sector species, i.e.\ $X \r \chi\chi$ and $X \r \psi {\psi}$. The DM relic abundance $\Omega_{\psi} h^2 = 0.12 \pm 0.0012$ \cite{ParticleDataGroup:2024cfk} is obtained once the interactions $\psi{\psi} \leftrightarrow \chi\chi$ freeze out.}
    \label{fig:mechanism}
\end{figure*}

In this work, we enlarge the dark sector to also contain a light ($\mathcal{O}(\rm keV)$) fermionic DM candidate $\psi$ with mass $m_\psi \gtrsim 10$~keV. In the case in which $m_X > 2m_\psi$, the $X$ boson produced at the expense of neutrinos will not only populate the dark radiation component but also produce the DM candidate via $X \r \psi\psi$, which thermalizes with the massless $\chi$ states and takes part in conversion processes such as $\psi{\psi} \leftrightarrow \chi\chi$, analogous to multi-component DM frameworks \cite{Belanger:2011ww,Liu:2011aa,BasiBeneito:2022qxd}. Alternatively, in the case where $m_X < m_\psi$, the DM can still be produced by $2 \lr 2$ interactions such as $\psi\psi \lr XX$ in addition to $\psi{\psi} \leftrightarrow \chi\chi$, which would establish an equilibrium in the dark sector. 

At temperatures below $m_\psi$, the $2 \leftrightarrow 2$ conversions between $\psi$, $\chi$ and $X$ will freeze-out, resulting in a slight increase in the $\chi$ population as well as determining the DM relic abundance. A schematic illustration of this mechanism is shown in fig.~\ref{fig:mechanism} for a chosen set of benchmark values, that allows us to relax the bound on the sum of neutrino masses by a factor $n_\nu/n_\nu^{\rm SM} \approx 0.2$. In the next section, we present a UV complete model to realize this mechanism. The parameter space of the scenario is rather constrained by various astrophysical and cosmological bounds, which we discuss in Sec.~\ref{sec:ps}. In general, dark sector--neutrino interactions such as $\nu\bar{\nu} \leftrightarrow X$ are potentially also subject to some constraints from the laboratory experiments, e.g., involving decays of SM particles such as mesons ($\pi^\pm ,K^\pm$), $\tau$ leptons and $Z_0, W^\pm$ gauge bosons, which we discuss at the end of Sec.~\ref{sec:model}.

\section{Model}
\label{sec:model}

For a minimal realization of the mechanism described above, we extend the SM by adding in total $N_{\rm heavy}$ heavy right-handed neutrinos (denoted by $N,N'$), involved in the generation of neutrino masses via the seesaw mechanism. A suitable choice for our purposes will be 3 copies of $N$ and one copy of $N'$, i.e., $N_{\rm heavy} = 4$.
Furthermore, we introduce a dark sector comprising a scalar singlet $\Phi$ and $N_{\rm light} = N_\chi + 1$ generations of light vector-like fermions (denoted by $\chi, \psi$). The dark sector particles $\psi,\chi$ and $\Phi$ are charged under a $U(1)_X$ symmetry, which can be either global or local. In ref.~\cite{Escudero:2022gez}, it was shown that the gauged version is more appealing than the global one, therefore, in the following we will consider the former case. See ref.~\cite{Ko:2014bka} for a different application of a similar model. 
In Table~\ref{tab:fields}, we list the new particles and their charge assignments. 

\begin{table}[t]
\small
    \centering
    \begin{tabular}{c c c c}
        \hline \hline
          & Field & Species & $U(1)_X$\\
          \hline
          \textbf{Scalar} & $\Phi$ & 1 & $+1$\\
          \hline
          \multirow{2}{*}{\textbf{Fermions}}& $\chi$ & $N_\chi$ & $-1$\\
          & $\psi$ & $1$ & $-1$\\
          & $N$ & 3 & 0\\
          & $N'$ & 1 & 0\\
          \hline \hline
    \end{tabular}
    \caption{List of particles added to the SM. We give the number of copies for each type of particle (``Species'') and their charge assignments.}
    \label{tab:fields}
\end{table}

Further, a $\mathbb{Z}_2$ symmetry is imposed, under which all particles except $\chi_R$ and $\psi_R$ are even, in order to avoid the vector-like mass terms and their interactions with $\Phi$ and $N$. (The $\chi_R$ and $\psi_R$ fields are needed only for anomaly cancellation.) Thus, the Lagrangian consistent with the charge assignments can be written as
\begin{align}\label{eq:lag}
    -\mathcal{L}_{\rm int} = &Y_\nu \bar{N} l_L \Tilde{H}^\dagger + Y_\chi \bar{N} \chi_L \Phi + Y_\psi \bar{N} \psi_L \Phi
    +Y'_\nu \bar{N'} l_L \Tilde{H}^\dagger + Y'_\chi \bar{N'} \chi_L \Phi + Y'_\psi \bar{N'} \psi_L \Phi\nonumber\\
    &+\frac{1}{2}M \bar{N} N^c + \frac{1}{2}{M'} \bar{N'} {N'}^c + {\rm H.c.}\,,
\end{align}
where $l_L$ and $H$ are the SM lepton and Higgs doublets respectively, $\Tilde{H}= i \sigma_2 H^\ast$,  and $M,M'$ are the right-handed neutrino masses of $N,N'$, respectively. The scalar potential can be written as
\begin{align}
    V(H,\phi) =& \mu_H^2 H^\dagger H + \lambda_H (H^\dagger H)^2 + \mu_\phi^2 \abs{\Phi}^2 +\lambda_\phi \abs{\Phi}^4 
    + \lambda_{H\phi} \abs{\Phi}^2 H^\dagger H\,,
\end{align}
where we take $\lambda_{H\phi}\approx 0$, to avoid the mixing between the scalars. After electroweak symmetry breaking, the SM Higgs takes a VEV which can be parameterized as $\bra H \ket = 1/\sqrt{2}\,[0\,, v_{\rm EW}]^T$ with $v_{\rm EW} =246$ GeV. Further, $\bra \Phi \ket = v_\phi/\sqrt{2}$ breaks the $U(1)_X$ symmetry, with $v_\phi^2 = -\mu_\phi^2/\lambda_\phi$, also generating the mass for the associated gauge boson $Z'$, $m_{Z'} = g v_\phi$, where $g$ is the $U(1)_X$ gauge coupling. \\


\subsection{Neutral fermion mixing}

The symmetry breaking induces mixing among the neutral leptons, and gives rise to the following 
$(3+N_{\rm light}+N_{\rm heavy}) \times (3+N_{\rm light}+N_{\rm heavy})$
mass matrix in the basis $(\chi_L^c,\nu_L^c,\psi_L^c,N',N)$:
\begin{equation}\label{eq:massmatrix}
\mathcal{M}_n = 
    \begin{pmatrix}
        0 & 0 & 0 & \Lambda' & \Lambda\\
        0 & 0 & 0 & {m_D}' & {m_D}\\
        0 & 0 & 0 & \kappa' & \kappa\\
        {\Lambda'}^T & {{m_D^T}'} & {\kappa'}^T & M' & 0\\
        {\Lambda}^T & {m_D^T} & {\kappa^T} & 0 & M
    \end{pmatrix}\,,
\end{equation}
where 
\begin{align}\label{eq:yukawas}
&m_D = Y_\nu v_{\rm EW}/\sqrt{2}\,,~{m_D}' = Y'_\nu v_{\rm EW}/\sqrt{2} \,,\nonumber \\ 
&\Lambda= Y_\chi v_\phi/\sqrt{2}\,,\qquad\Lambda'= Y'_\chi v_\phi/\sqrt{2}\,, \nonumber \\ 
&\kappa = Y_\psi v_\phi/\sqrt{2}\,,\qquad\kappa'= Y'_\psi v_\phi/\sqrt{2}\,. 
\end{align}
The rank of the matrix \eqref{eq:massmatrix} is $2N_{\rm heavy}$, leading to $(3+N_{\rm light}-N_{\rm heavy})$ massless and $2N_{\rm heavy}$ massive states. For our purposes, we want 4 massive states in addition to the $N_{\rm heavy}$ heavy right-handed neutrinos: the 3 active neutrinos plus the DM candidate. Therefore, we chose $N_{\rm heavy} = 4$, and $N_{\rm light} = N_\chi + 1$, leaving $N_\chi$ states massless. One of the ``light'' dark sector fermions gets massive, which will become our DM candidate, and we single it out by denoting it with $\psi$ to distinguish it from its massless partners $\chi$.\footnote{Here we assume that all three active neutrinos are massive. If the lightest of them remains massless, we would need only $N_{\rm heavy} = 3$ heavy right-handed neutrinos.}

Note that only left-handed fields appear in the Yukawa Lagrangian eq.~\eqref{eq:lag} and receive masses according to \eqref{eq:massmatrix}, whereas the right-handed fields $\chi_R$ and $\psi_R$ remain massless due to the postulated $\mathbb{Z}_2$ symmetry. Hence, we are left with
\begin{equation} \label{eq:massless}   
\tilde{N} = 2N_\chi + 1\,,~\tilde g = 4N_\chi + 2
\end{equation}
massless states and degrees of freedom in total, respectively, corresponding to $\chi_L,\chi_R$ and $\psi_R$.

Upon block diagonalisation (see Appendix~\ref{sec:diag} for details) we get the following mass eigenvalues
\begin{align}\label{eq:eigenvalues}
    &m_\chi = 0\,, \nonumber\\
    &m_{\nu} = \frac{(m_D \kappa' - {m_D}' \kappa)^2 + ({m_D}'\Lambda-m_D \Lambda')^2 + (\kappa'\Lambda-\kappa\Lambda')^2}{M'(m_D^2 + \kappa^2 + \Lambda^2)+M({m_D}'^2+\kappa'^2+\Lambda'^2)}\,, \nonumber\\
    &m_{\psi} \approx  \frac{m_D^2 + \kappa^2 +\Lambda^2}{M}+\frac{{m_D}'^2+\kappa'^2+\Lambda'^2}{M'}\,, \nonumber\\
    &m_{N'} \approx  M'\,, \nonumber\\
    &m_{N} \approx  M  \,.
\end{align}
Here, $m_{\nu} = U_\nu^\ast \hat{m}_\nu U_\nu^\dagger$, where $U_\nu$ is the PMNS mixing matrix in the diagonal mass basis for the charged leptons and $\hat{m}_\nu = {\rm diag}(m_1,m_2,m_3)$ contains the physical neutrino mass eigenvalues, and we use $a^2 = a a^T$ in order to write the equations for $m_{\nu,\psi}$ in a compact form.

The mass eigenstates are then obtained as
\begin{align}
    (\hat{\chi},\hat{\nu},\hat{\psi},\hat{N'},\hat{N})^T = \mathbf{W}^\dagger ({\chi_L^c},{\nu_L^c},{\psi_L^c},{N'},{N})^T\,,
\end{align}
where the mixing matrix $\mathbf{W}$ is given by
\begin{align}\label{eq:mixingmat}
\mathbf{W}=&
    \begin{pmatrix}
        1 & \frac{\Lambda^\ast}{m_D^\ast} & 0 & \frac{{\Lambda'}^\ast}{{M'}^\dagger} & \frac{\Lambda^\ast}{M^\dagger}\\
        \frac{-\Lambda^T}{m_D^T} & 1 & \frac{{m_D}'^\ast}{\kappa'^\ast} & \frac{{m_D}'^\ast}{M'^\dagger} & \frac{m_D^\ast}{M^\dagger}\\
        0 & \frac{-{m_D}'^T}{\kappa'^T} & 1 & \frac{\kappa'}{M'} & \frac{\kappa}{M}\\
        {\frac{{-\Lambda}'^T}{M'}} & \frac{-{m_D}'^T}{M'} & \frac{-\kappa'^T}{M'} & 1 & 0\\
        \frac{-\Lambda^T}{M} & \frac{-m_D^T}{M} & \frac{-\kappa^T}{M} & 0 & 1
    \end{pmatrix}\times {\rm~Diag}[1,U_\nu,1,1,1]\,.
\end{align}

In deriving the mixing matrix above, we adopt a diagonal basis for the right-handed neutrino mass matrix, and assume the following hierarchy: 
\begin{align}\label{eq:hierarchy}
    M \gg M' \gg m_D \gg \kappa', \Lambda \gg {m_D'}, \Lambda',\kappa \,.
\end{align}

This corresponds to a regime where the dominant interactions of $N'$ are with $\psi$ whereas $N$ interacts dominantly with $\chi$ and $\nu$. Indeed, in the limit of $m_D', \Lambda' ,\kappa \to 0$, the mass matrix becomes block-diagonal and the $\psi-N'$ sub-block decouples. In this limit, we recover an additional 
$\mathbb{Z}_2$ symmetry under which $\psi_L$ and $N'$ are odd and all other fields even. Hence, the smallness of $m_D', \Lambda',\kappa$ in eq.~\eqref{eq:hierarchy} can be motivated by this approximate
$\mathbb{Z}_2$ symmetry. Indeed, for the phenomenology discussed below, including DM production, we do not need these couplings and the symmetry could even be exact, enforcing $m_D'  = \Lambda'= \kappa = 0$. Note that given the assumed hierarchy, all the mixing terms are quite small, thus, contributions from oscillations amongst the light fields can be safely neglected. 

Using the hierarchy above, we can simplify the expressions in eq.~\eqref{eq:eigenvalues} for the neutrino and dark matter mass
\begin{align}
    &m_\nu \approx \frac{m_D^2{\kappa'}^2}{M'm_D^2 + M{\kappa'}^2} \,, 
    \nonumber\\
    &m_\psi \approx \frac{m_D^2}{M} + \frac{\kappa'^2}{M'}\,.
\end{align}
We see that $m_\psi$ receives two seesaw contributions, from both, $N$ and $N'$. In order to have $\psi$ as viable DM particle we need $m_\psi \gg m_\nu$, which can be achieved by imposing an additional condition, namely 
\begin{equation}\label{eq:hier2}
M'm_D^2 \ll M {\kappa'}^2 \,,    
\end{equation}
which does not follow from the hierarchy \eqref{eq:hierarchy} but is compatible with it.\footnote{We can achieve eq.~\eqref{eq:hier2} by requiring $M \ggg M'$. For instance, for large Yukawa couplings typically we will have ${m_D} \sim 100$~GeV and $\kappa' \sim 1$~GeV, so we need $M'/M \ll 10^{-4}$.} Then we obtain
\begin{align}\label{eq:nudmmass}
    &m_{\nu} \approx m_D M^{-1} m_D^T\,, \nonumber\\
    &m_\psi \approx \kappa' M'^{-1} \kappa'^T\,.
\end{align}
Hence, the seesaw mechanism factorizes, such that the three $N$ are responsible to generate active neutrino masses, whereas $N'$ is responsible for the DM mass $m_\psi$. (This motivates our notation, to distinguish $N$ and $N'$.)

Adopting the one-flavor approximation, the above mixing matrix can be expressed in terms of the mixing angles $\theta$ at the leading order
\begin{align}
    \theta_{\nu N} &= \frac{m_D}{M}\,,\quad \theta_{\nu \chi} = \frac{\Lambda}{m_D}\,,\quad \theta_{N \chi} = \frac{\Lambda}{M}\sim 0\,,\nonumber\\
    \theta_{\nu N'} &= \frac{{m_D}'}{M'}\,,\quad \theta_{\nu \psi} = \frac{{m_D}'}{\kappa'}\,,\quad \theta_{N' \psi} = \frac{\kappa'}{M'}\,.
\end{align}

\subsection{Parameters}

Here, we discuss the parameters and interactions relevant for phenomenology. Using the seesaw relation in eq.~\eqref{eq:nudmmass}, we can express $\theta_{\nu N}$ in terms of the neutrino mass as $\theta_{\nu N} = \sqrt{m_{\nu}/M}$, and similarly $\theta_{N' \psi} = \sqrt{m_{\psi}/M'}$. It can also be seen that $\theta_{\nu N'} = \theta_{\nu \psi} \theta_{N' \psi}$, hence, for our analysis, we are left with the following independent parameters
\begin{equation*}
    \{m_{\nu}, m_\psi, M', M, \theta_{\nu \chi}, \theta_{\nu \psi},v_\phi,m_{Z'},N_\chi\}\,.
\end{equation*}
We fix $m_\nu =0.2$ eV for numerical estimates and are interested in the DM mass $m_\psi \sim \mathcal{O}(10-100)$ keV. For instance,
\begin{equation}\label{eq:dm_mass}
    m_\psi \sim 10 {\rm~keV}\,\left(\frac{\kappa'}{10^4{\rm~keV}}\right)^2\left(\frac{10{\rm~GeV}}{M'}\right)\,,
\end{equation}
tells us about the order of magnitude of $M'$ and $\kappa' = Y'_\psi v_\phi$. Note that the value of $M$ is not relevant for the mechanism described above, however, it is constrained by perturbativity requirements and further constraints arise if we also want to incorporate thermal leptogenesis in the model \cite{Escudero:2022gez}. 

For simplicity, we work in the one-flavor approximation and assume that the interactions between active neutrinos and all $N_\chi$ species have the same strength parameterized by $\theta_{\nu\chi}$. The mixing angle $\theta_{\nu\psi}$, instead, determines the interaction between neutrinos and DM which is subject to several constraints from DM stability and $X$-ray observations, as we discuss later. Hence, in order to evade these constraints, this mixing must be highly suppressed, i.e., $\theta_{\nu\chi} \gg \theta_{\nu\psi}$. In the limit of the above mentioned $\mathbb{Z}_2$ symmetry, we would even have $m_D' = 0$ and therefore $\theta_{\nu\psi} =0$. 
Hence, to determine the viable parameter space of the model, the main parameters of interest are: 
\begin{equation}\label{eq:param}
\{m_\psi, m_{Z'}, v_\phi, \theta_{\nu \chi},N_\chi\} \,.    
\end{equation}

\subsection{Interactions}
 
Apart from the Yukawa interactions that can be directly read from the Lagrangian in eq.~\eqref{eq:lag}, we will also have gauge boson-mediated interactions. The interactions of the fermions with the $U(1)_X$ gauge boson can be written as
\begin{equation}
    \mathcal{L} = \sum_{f} Q_f g {Z'}_\mu \bar{f}\gamma^\mu f\,,
    \label{eq:Zprime}
\end{equation}
where $f = \{\chi_L,\chi_R, \psi_L, \psi_R\}$, 
\begin{equation}
g \equiv \frac{m_{Z'}}{v_\phi}    
\end{equation}
is the $U(1)_X$ gauge coupling, $Q_f = -1$ denotes the $U(1)_X$ charge for both the left and right-handed fermions. Given the mixing among the active and sterile neutrinos defined above, this interaction will give rise to the couplings $\lambda_{Z'}^{xx}$ leading to processes $Z' \leftrightarrow xx$
\begin{align}
    &\lambda_{Z'}^{\nu \nu} = \frac{m_{Z'}}{v_\phi}(\theta_{\nu \chi}^2 + \theta_{\nu \psi}^2) \simeq \frac{m_{Z'}}{v_\phi} \theta_{\nu\chi}^2\,,\nonumber\\    
    &\lambda_{Z'}^{\nu \chi} = \frac{m_{Z'}}{v_\phi}\theta_{\nu \chi} \,,\nonumber\\
    &\lambda_{Z'}^{\nu \psi} = \frac{m_{Z'}}{v_\phi}\theta_{\nu \psi} \,,\nonumber\\
    &\lambda_{Z'}^{\psi \psi} = \lambda_{Z'}^{\chi \chi} = \frac{m_{Z'}}{v_\phi} \,.
    \label{eq:lambda}
\end{align}
For our purposes, these gauge interactions are dominant compared to the real and pseudoscalars interactions between $\Phi$ and $\nu,\chi$ which is further suppressed by $m_\nu$. Therefore, we neglect the latter here. The couplings above will lead to various decays, inverse decays, and $2 \leftrightarrow 2$ processes, the rates of which are collected in Appendix~\ref{sec:rates}. 

We briefly comment on laboratory constraints from SM particle decays involving neutrinos, such as $Z_0,W^\pm,K^\pm, \pi^\pm$, which in principle are sensitive to neutrino--$Z'$ or neutrino--(pseudo-)scalar
interactions, as the emission of a mediator from the neutrino leg would lead to observable modifications of the decays, implying strong bounds on the corresponding coupling, see \cite{Laha:2013xua,Bakhti:2017jhm,Arcadi:2018xdd,Kelly:2019wow,Brdar:2020nbj,Dev:2024twk}. These bounds do not apply to our model, as all light mass states mix with the active neutrinos and $\chi$'s contribute to the internal neutrino propagator. Therefore, the emission of the $Z'$ is highly suppressed due to the approximate unitarity of the $\nu-\chi$ sub-block of the mixing matrix (which is unitary up to tiny corrections due to mixing with heavy states, $\theta_{\nu N},\theta_{\nu N'}$). The same argument can be applied to interactions with the (pseudo-) scalar bosons, which in addition are further suppressed by a factor $m_\nu/m_{Z'}$ relative to $Z'$ interactions. Therefore, such laboratory bounds do not lead to additional constraints on the parameter space of our model on top of the astrophysical and cosmological limits discussed in sec.~\ref{sec:ps} below.

\section{Dark Matter}\label{sec:dm}

\begin{figure}[!tb]
    \centering
    \includegraphics[scale=0.4]{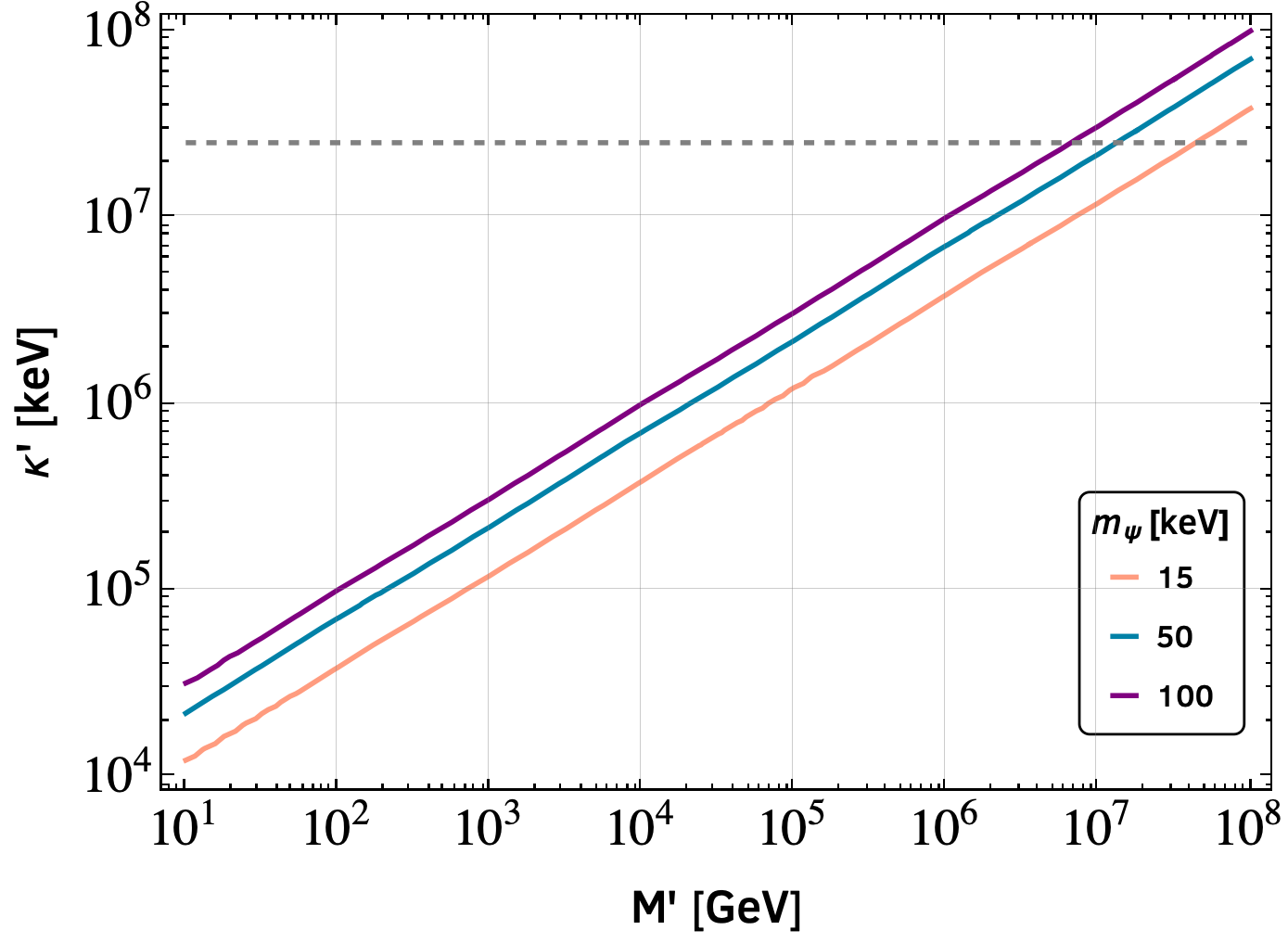}
    \caption{Contours of fixed DM mass $m_\psi = \{5,15,50,100\}$~keV in the plane of $N'$ mass $M'$ and $\kappa' = Y'_\psi v_\phi/\sqrt{2}$. The gray dashed line indicates the bound from perturbativity of the Yukawa coupling $Y'_\psi$, taking $v_\phi^{\rm max} \sim 10$ GeV.}
    \label{fig:dmmass}
\end{figure}

The DM candidate gets its mass via the seesaw mechanism similar to the active neutrinos (see eq.~\eqref{eq:nudmmass}) while the $\chi$'s remain massless. In fig.~\ref{fig:dmmass}, we show the contours of fixed DM mass for different values of $M'$ and $\kappa'$. We also indicate the maximum bound on $\kappa'$ from the perturbativity of the Yukawa coupling ${\kappa'}^{\rm max} \simeq \sqrt{4\pi}v_\phi^{\rm max}/\sqrt{2}$ by the dashed gray line. We take $v_\phi^{\rm max} \sim 10$ GeV, after considering several constraints that are discussed in the next section. 
Finally, we notice that considering $m_\psi \sim \mathcal{O}(10-100)$ keV corresponds to the upper bound on $M' \lesssim 10^8$ GeV. Below, we discuss the DM production, freeze-out and constraints from relic abundance, stability and structure formation.

\subsection{Production and freeze-out}\label{sec:prod}

At early times, DM may be produced from the decays of heavy neutrinos $N, N'$ via the Yukawa interactions in eq.~\eqref{eq:lag}, while the production from $\nu \leftrightarrow \psi$ oscillations is negligible due to $\theta_{\nu\psi} \ll 1$. Later, once we have a significant population of $Z'$ bosons, $\psi$ is produced dominantly from $Z'$ decays for $m_{Z'} > 2 m_\psi$, or via processes such as 
$Z' Z' \leftrightarrow \psi \psi$ and $\chi\chi \leftrightarrow \psi\psi$ when $m_{Z'} < 2 m_\psi$. In order to not spoil successful BBN, we need to make sure that these processes do not bring $\psi$ or $\chi$ in equilibrium with the SM plasma before neutron-proton freeze-out. 

To start, let's consider the DM production via heavy neutrino decays. Imposing the out-of-equilibrium condition $\Gamma < H$ gives $Y < 10^{-8} ({M}/{{\rm GeV}})^{1/2}$. Since the Yukawa interactions of $N$ with the SM lepton and Higgs doublet determine the $\nu$ masses, they are sizeable and expected to thermalize $N$ with the SM plasma. In order to not bring $\psi$ into equilibrium with the SM, we require that $\kappa$ is sufficiently small--consistent with the hierarchy \eqref{eq:hierarchy}, such that $N$ does not play a significant role in DM production. Furthermore, in some regions of the parameter space we have to assume a re-heating temperature $T_{\rm RH} < M_N$, such that $N$ are never produced, in order to avoid early equilibration of massless $\chi$ from $N$ decays \cite{Escudero:2022gez}. 

On the other hand, with the hierarchies (\ref{eq:hierarchy}, \ref{eq:hier2}), for $N'$
the relation $Y'_\nu < 10^{-8} (\frac{M'}{{\rm GeV}})^{1/2}$ is fulfilled and therefore $N'$ does not thermalize with the SM plasma. Thus also $N'$ decays do not lead to a significant production of $\psi$, despite the sizeable coupling $\kappa'$ required to generate the DM mass, see eq.~\eqref{eq:dm_mass} and fig.~\ref{fig:dmmass}.

At later times, the decays and scatterings involving $Z'$ populate both $\psi$ and $\chi$, establishing an equilibrium in the dark sector, including also SM neutrinos. The equilibration happens after neutrino decoupling from the SM, therefore leading to a suppressed neutrino number density. Once $T < m_\psi$, the annihilations $\psi\psi \r \chi\chi$ and $\psi\psi \r Z' Z'$ (for $m_{Z'}<m_\psi$) will deplete the $\psi$ population,\footnote{Note that conversions such as $\psi\psi \r \nu\nu {\rm~or~} \nu\chi$ are also present, however, they are suppressed by the mixing angle $\theta_{\nu\chi}$ and subdominant enough to be safely neglected in determining the DM relic abundance.} eventually leading to their freeze-out and determining the DM relic abundance. This is analogous to the WIMP freeze-out, but in the dark sector characterized by a temperature $T_{\rm dark}$ with the DM freezing-out at a temperature of the order of a few keVs. 
This DM production mechanism is similar to the scenario discussed in refs.~\cite{Berlin:2017ftj,Berlin:2018ztp}, extended by the massless fermions in our model; see also refs.~\cite{Feng:2008mu, Berlin:2016gtr} for  discussions of DM freeze-out in a hidden sector.

\begin{table}[t]
\small
    \centering
    \begin{tabular}{C{1.8cm} C{1.8cm} C{2.8cm} C{1.8cm}}
        \hline \hline
          & $m_{Z'}>2m_\psi$ & $2m_\psi>m_{Z'}> m_\psi$ & $m_{Z'}<m_\psi$\\
          \hline
          \multirow{2}{*}{\textbf{Production}}& $Z' \lr \psi\psi$ & $Z'Z' \lr \psi\psi$ & $Z'Z' \lr \psi\psi$\\
          & & $\chi\chi \lr \psi\psi$ & $\chi\chi \lr \psi\psi$\\
          \hline
          \multirow{2}{*}{\textbf{Depletion}}& $\psi\psi \r \chi\chi$ & $\psi\psi \r \chi\chi$ & $\psi\psi \r \chi\chi$\\
          & & & $\psi\psi \r Z'Z'$ \\
          \hline
          \multirow{2}{*}{\textbf{Decay}}& $\psi \r \nu\chi\chi$ & $\psi \r \nu\chi\chi$ & $\psi \r \nu\chi\chi$\\
          & & & $\psi \r Z'\nu$ \\
          \hline \hline
    \end{tabular}
    \caption{Processes contributing to the production, freeze-out and decay of DM candidate $\psi$ in our model for different kinematical regimes.}
    \label{tab:regime}
\end{table}

In Table~\ref{tab:regime}, we summarize the dominant production and depletion channels ($\sigma \propto g^4,\Gamma \propto g^2$) for different kinematical regimes for $m_{Z'}$ and $m_\psi$. We also list the possible modes of DM decay ($\Gamma \propto \theta_{\nu\psi}^2$) in each regime. In order to track the evolution of number densities of different BSM species in presence of the aforementioned processes, we numerically solve a set of coupled Boltzmann equations collected in Appendix~\ref{sec:rates}. A representative example is shown in fig.~\ref{fig:mechanism} for the case $m_{Z'} > 2 m_\psi$. 

\subsection{Relic abundance}

The observed DM relic abundance $\Omega_{\psi} h^2 = 0.12 \pm 0.0012$ \cite{ParticleDataGroup:2024cfk} is obtained once the interactions between $\psi$, $\chi$ and $Z'$ freeze-out. The dominant DM annihilation rates are given by (see e.g., refs. \cite{Berlin:2014tja,Arcadi:2024ukq})
\begin{align}\label{eq:dmann}
     (\sigma v)_{\psi\psi\to\chi\chi} &\approx \tilde{N}\, \frac{g^4}{48\pi}\frac{m_\psi^2}{(m_{Z'}^2-4m_\psi^2)^2}v^2\,,\nonumber\\
    (\sigma v)_{\psi\psi\to Z' Z'} &\approx \frac{g^4}{16\pi\, m_\psi^2}\left({1-\frac{m_{Z'}^2}{m_\psi^2}}\right)^{1/2}\left(1+\frac{m_\psi^4}{m_{Z'}^4}v^2 + \mathcal{O}\left(\frac{m_{Z'}^2}{m_\psi^2}\right)\right)\,,
\end{align}
and the thermal average can be obtained by substituting $\braket{v^2}=6/x_d$ with $x_d \equiv m_\psi/T_{\rm dark}$. Here we take into account that the dark sector temperature $T_{\rm dark}$ will deviate from the photon temperature $T_\gamma$ in our model, see discussion in section~\ref{sec:neff}.
For $m_{Z'} \lesssim m_\psi$, quickly the second term in the $\psi\psi\to Z'Z'$ cross-section dominates over the others. It can be seen that for $m_\psi < m_{Z'}$ the number of massless species $\tilde{N}$ plays an important role in determining the strength of $\psi\psi \r \chi\chi$ annihilations in addition to relaxing the cosmological mass bound.\footnote{When not specifying chirality, we use a shorthand notation with $\psi$ referring actually to the massive $\psi_L$, whereas $\chi$ collectively denotes the massless states $\chi_L,\chi_R,\psi_R$, compare eq.~\eqref{eq:massless}.}
The DM relic abundance is computed as
\begin{equation}
    \Omega_\psi h^2 \approx 2.755 \times 10^2 \, \left(\frac{m_\psi}{1 {\rm~keV}}\right) Y_\psi^{\infty}\,,
\end{equation}
where $Y_\psi^{\infty}$ is the asymptotic DM yield defined in appendix~\ref{sec:rates} and obtained by solving eq.~\eqref{eq:be} numerically. 

Analogous to the visible sector freeze-out, we can use the analytical approximation for the relic abundance \cite{Kolb:1990vq}. It is convenient to express the dark sector temperature in terms of the visible sector temperature by defining $\xi \equiv T_{\rm dark}/T_{\gamma}$, so we have 
\begin{align}\label{eq:analytical}
    \Omega_\psi h^2 \simeq x_f \frac{10^{-10} \, {\rm GeV}^{-2}}{\braket{\sigma v}_{\rm tot}}\,,
\end{align}
with $\braket{\sigma v}_{\rm tot} = \bra \sigma v \ket_{\psi\psi\to\chi\chi}+\bra \sigma v \ket_{\psi\psi\to Z' Z'}$. Here, $x_f = m_\psi/T_{\gamma,f}$ denotes the freeze-out temperature from the visible sector viewpoint, and it depends on $\xi$ as \cite{Feng:2008mu}
\begin{align}\label{eq:xfan}
    x_f &\approx \xi \ln[0.04 \delta(\delta+2)(g_\psi/g_{\rm eff}^{1/2}) \beta\xi^{3/2}]-\frac{1}{2}\xi\ln\{\xi \ln[0.04 \delta(\delta+2)(g_\psi/g_{\rm eff}^{1/2}) \beta\xi^{3/2}]\}\,,
\end{align}
where $g_\psi=2$ are the internal degrees of freedom of $\psi$, $g_{\rm eff} = g_\gamma + g_{\rm dark}\xi^4$ takes into account the contribution from the total relativistic degrees of freedom in the dark sector, $\delta$ is a numerical factor chosen to match the numerical solutions (we take $\delta \approx 0.7$ for $p$-wave annihilations), and $\beta = M_{\rm pl}\, m_\psi\, \braket{\sigma v}_{{\rm tot}}^{x=1}$ with $M_{\rm pl} = 1.2 \times 10^{19}$ GeV. In our scenario, the dark sector temperature varies with the number of massless species $\tilde{N}$, as we will discuss in sec.~\ref{sec:neff}, and therefore, $x_f$ also depends on $\tilde{N}$ via the temperature ratio $\xi$. Using eq.~\eqref{eq:xfan}, we find that DM freeze-out occurs at $x_f \sim 3-5$ for the relevant temperature window and annihilation cross-sections (in agreement with the numerical solution of the Boltzmann equations).
\begin{figure}[!tb]
    \centering
    \includegraphics[scale=0.4]{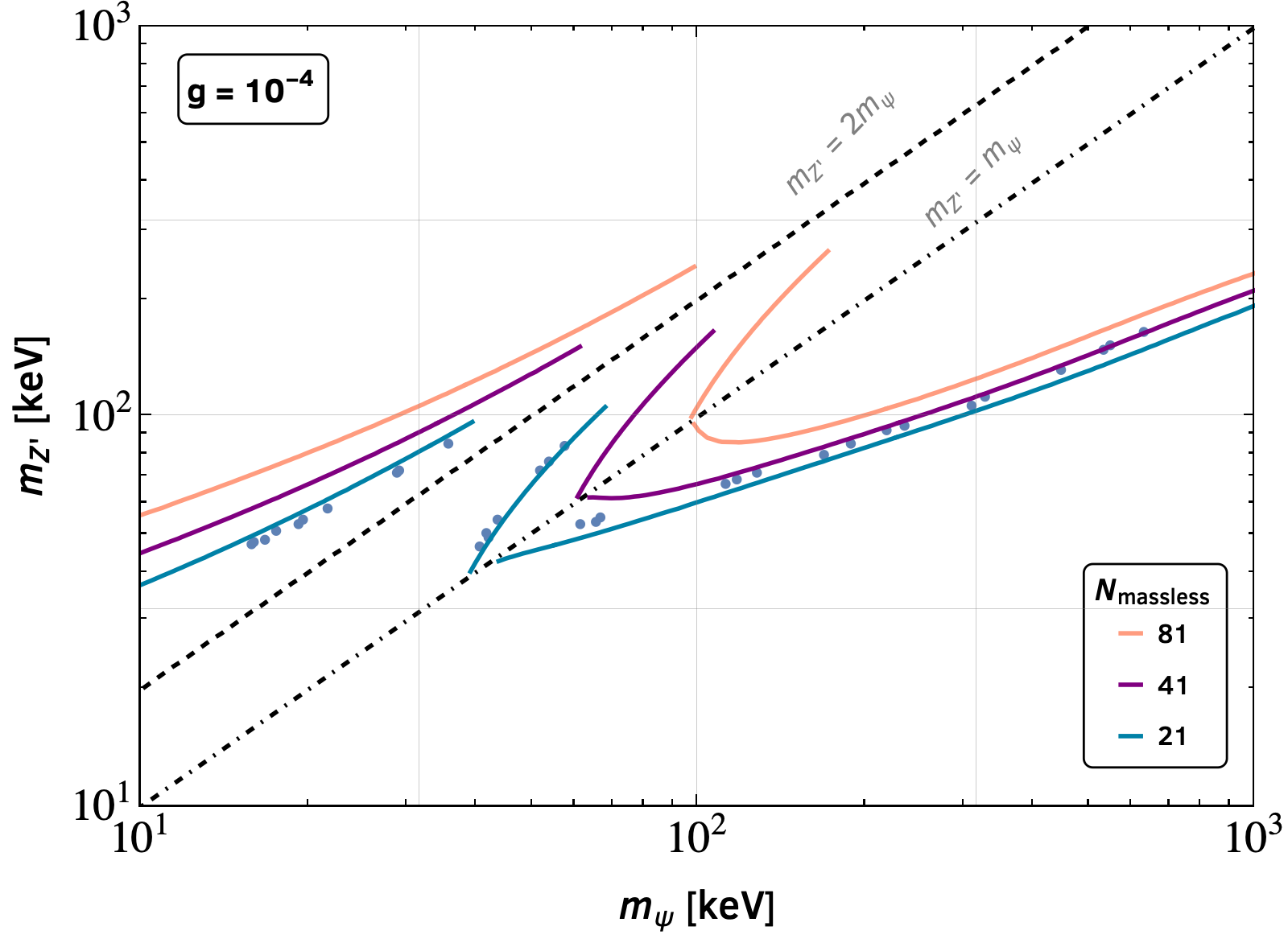}
    \caption{Contours of observed DM relic abundance for $g=10^{-4}$ in the $m_{Z'}-m_\psi$ plane for different massless degrees of freedom, obtained from the analytical approximation of eq.~\eqref{eq:analytical}. The blue dots represent the numerical solution for $\tilde{N} =21$, matching closely with the analytical one. The dashed black line represents the $s$-channel resonance condition $m_{Z'} = 2m_{\psi}$, whereas the dash-dotted line corresponds to $m_{Z'}=m_\psi$, below which the  $t$-channel annhilations $\psi\psi\to Z'Z'$ dominate.}
    \label{fig:dmom}
\end{figure}
In fig.~\ref{fig:dmom}, we show the contours of observed relic abundance for a fixed value of the gauge coupling ($g=10^{-4}$) and for varying $\tilde{N}$ in the $m_{Z'}-m_\psi$ plane using the analytical solution given above. For the case of $\tilde{N}=21$, we also show the numerical result by solving the Boltzmann equations by blue points, matching closely the blue curve corresponding to the analytical approximation. For $m_Z' > m_{\psi}$, where the annihilation to $\chi$ dominates, the dependence on $\tilde{N}$ follows explicitly from eq.~\eqref{eq:dmann}. In the region where annihilation to $Z'$ dominates, i.e., when $m_\psi$ is much larger than $m_{Z'}$, the $\tilde N$ dependence appears due to the modification of $x_f$ via $\xi$. It should be noted that near the resonance region, i.e., when $m_{Z'} \simeq 2 m_\psi$, the analytical solution and the thermally averaged cross-sections (eq.~\eqref{eq:dmann}) are not a good approximation, and a careful treatment of the cross-section and the freeze-out temperature is required in order to properly determine the DM relic abundance. Therefore, for dealing with resonances, we use the complete formula for thermal averaging (see Appendix~\ref{sec:rates}) and compute the relic abundance numerically after solving the Boltzmann equations. We find that in the region very close to the resonance, the annihilation rate during freeze-out is so large that we do not obtain the correct abundance.   

\begin{figure}[!tb]
    \centering
    \includegraphics[width=0.48\linewidth]{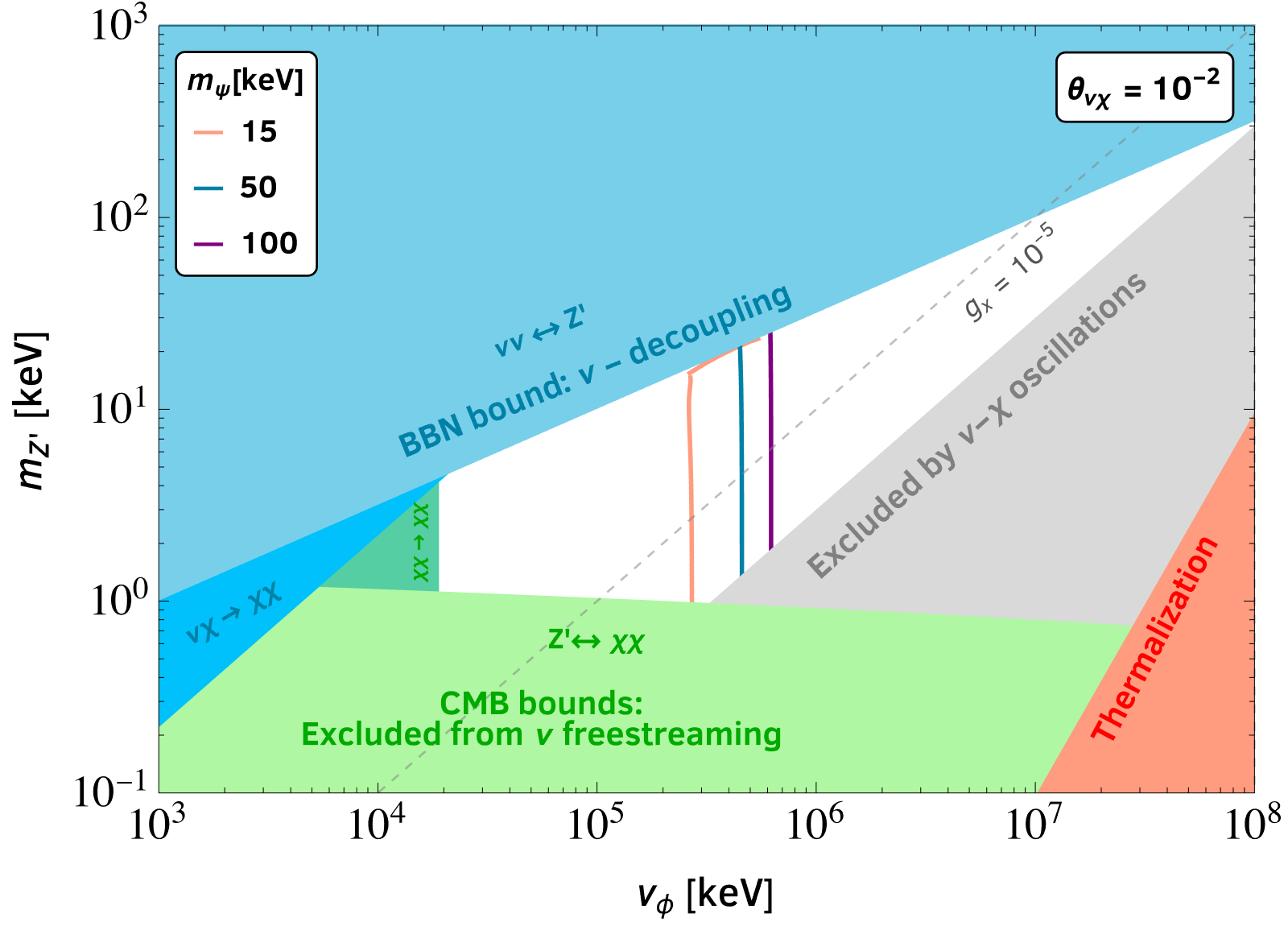}
    \includegraphics[width=0.48\linewidth]{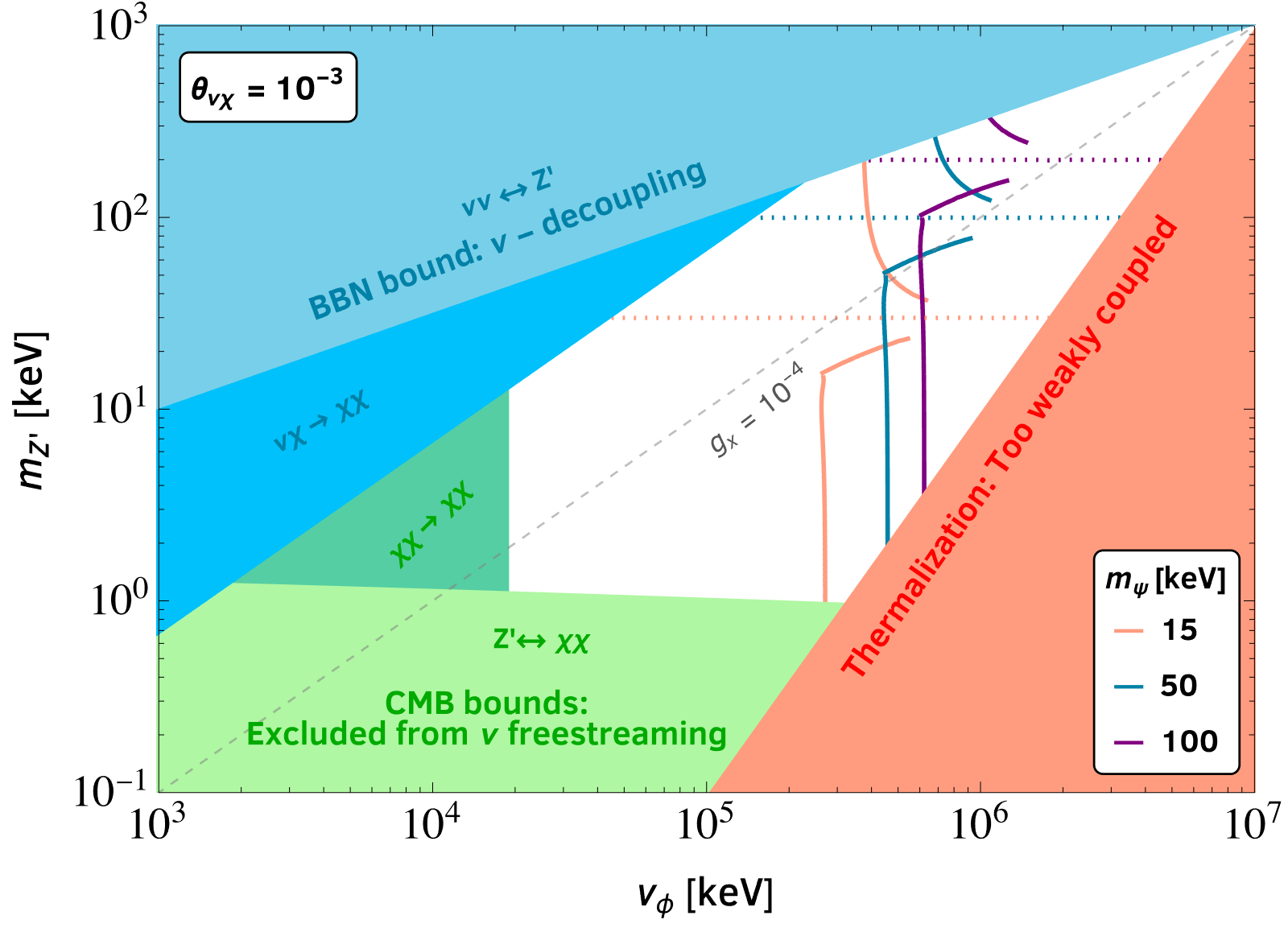}
    \caption{Parameter space of the model with the shaded areas highlighting regions of the parameter space excluded by several cosmological constraints, for a fixed value of $\nu-\chi$ mixing $\theta_{\nu\chi}=10^{-2}$ (\textit{left}) and $\theta_{\nu\chi}=10^{-3}$ (\textit{right}), and $N_\chi = 10$. Along the orange, blue, purple solid curves the observed relic DM density is obtained for $m_\psi = 15,50,100$~keV, respectively. The gray dashed lines indicate a fixed value of the $U(1)_X$ gauge coupling, $g = m_{Z'}/v_\phi$ and the dotted lines correspond to $m_{Z'} =2 m_\psi$ for a given DM mass. The red region is excluded from the thermalization condition, as the interactions of $Z'$ with neutrinos are not strong enough. The blue regions are excluded from BBN by requiring that $Z'$ is not in equilibrium with $\nu$'s at $T > 0.7$ MeV, and the green regions show the area excluded by $\nu$-free-streaming and CMB power spectra. 
    The gray shaded region is excluded from production of $\chi$ via $\nu-\chi$ oscillations before BBN.}
    \label{fig:ps}
\end{figure}

In fig.~\ref{fig:ps}, we show with solid lines of different colors contours of different DM mass which satisfy the relic abundance, highlighting the parameter space compatible with accommodating keV scale DM in the model as well  as relaxing the cosmological neutrino mass bound. This is the main result of our work. 

Note that in the limits far away from the resonance, the annihilation cross-sections from eq.~\eqref{eq:dmann} depend only on the combination $g/m_{Z'} = v_\phi^{-1}$ in both limits:
\begin{align}\label{eq:XSlimits}
    \langle \sigma v\rangle = \frac{m_\psi^2}{8\pi v_\phi^4 x_d} \times
    \left\{ 
    \begin{array}{l@{\quad}l}
       \tilde{N} &  (m_{Z'}\gg m_\psi)\\
       3  & (m_{Z'}\ll m_\psi)
    \end{array}
    \right. \,,
\end{align}
where we have used $\langle v^2\rangle = 6/x_d$
Hence, in these limits, the DM relic abundance according to eq.~\eqref{eq:analytical} fixes the dark VEV for a given DM mass:
\begin{align}\label{eq:v_phi}
    v_\phi \simeq & ~10^5\,{\rm keV} \left(\frac{m_\psi}{15\,{\rm keV}}\right)^{1/2}\left(\frac{3.2}{x_f}\right)^{1/2}\times
    \left\{ 
    \begin{array}{ll}
       2 \tilde{N}^{1/4}  &  (m_{Z'}\gg m_\psi)\\
       2.4  & (m_{Z'}\ll m_\psi)
    \end{array}
    \right. \,,
\end{align}
in agreement with fig.~\ref{fig:ps}. For large mixing angles $\theta_{\nu\chi}$ (upper panel of fig.~\ref{fig:ps}), we are in the limit $m_{Z'} < m_\psi$ in large regions of the parameter space and therefore the DM contours are determined by the combination $m_\psi / v_\phi^2 = g^2 m_\psi / m_{Z'}^2$, and depend mildly on $\tilde N$ due to $x_f$.

\subsection{Stability and X-ray constraints}

In our model, the interaction of $\psi_L$ and SM neutrinos with the massless states $\chi_{L,R}, \psi_R$ mediated by the gauge boson $Z'$ can lead to DM decay via three body processes. The dominant ones are $\psi \r \nu \chi\chi$.\footnote{The decay $\psi \r 3\nu$ is also allowed, but the amplitude of this process is suppressed by the $3^\text{rd}$ power of the mixing angle $\theta_{\nu\chi}$, instead of the single power of $\theta_{\nu\chi}$ that suppresses $\psi \r \nu\chi\chi$. Therefore, to account for DM stability, we only need to consider $\psi \r \nu\chi\chi$. Moreover, due to the absence of $Z' \psi\chi$ interactions, decays such as $\psi \r 3\chi$ do not occur.} Therefore, in order for the DM to be stable on cosmological timescales, we need to ensure that the lifetime of $\psi$ relative to this channel is at least larger than the age of the universe, $t_\text{U} \approx 4.35 \times 10^{17}{\rm~s}$. Using eq.~\eqref{eq:dmdecay} and trading the dependence on $g$ and $m_{Z'}$ by $v_{\phi}$ we get
\begin{align}\label{eq:dms}
    \tau_\psi \sim 45\, \tau_U \frac{\left(\frac{v_\phi}{2  {\rm~GeV}}\right)^4}{\left(\frac{\tilde{N}}{21}\right)\,\left(\frac{m_\psi}{15 {\rm~keV}}\right)^5 \,\left(\frac{\theta_{\nu\psi}}{10^{-8}}\right)^2}\,,
\end{align}
To be safe from cosmological constraints on the DM lifetime \cite{Audren:2014bca,Poulin:2016nat}, we impose $\tau_\psi > 20 \, \tau_U$, which gives
\begin{equation} \label{eq:lifetime-bound}
    \theta_{\nu\psi}^2 < 2 \times 10^{-16}
\left(\frac{15 {\rm~keV}}{m_\psi}\right)^5 \, 
    \left(\frac{21}{\tilde{N}}\right)\,
    \left(\frac{v_\phi}{2 {\rm~GeV}}\right)^4\,.
\end{equation}
We see that the DM lifetime bound sets a strong limit on the DM--neutrino mixing. We also observe preference for large values of $v_\phi$ in order to ensure the stability for $\mathcal{O}(10){\rm~keV}$ DM. 

On the other hand, if $m_\psi > m_{Z'}$, the $\psi \r Z'\nu$ decay channel opens up. Since it is a two-body decay it is comparatively unsuppressed with respect to the three-body decay discussed above (see eq.~\eqref{eq:dmdecay}) and the relative DM lifetime is given by 
\begin{equation}
    \tau_\psi \equiv \frac{1}{\Gamma_{\psi \r Z' \nu}} \sim 25 \tau_U \frac{\left(\frac{m_{Z'}}{10 {\rm~keV}}\right)^2}{\left(\frac{g}{10^{-4}}\right)^2\left(\frac{\theta_{\nu\psi}}{10^{-15}}\right)^2\left(\frac{m_\psi}{40 {\rm~keV}}\right)^3}\,,
\end{equation}
where the approximation holds well for $m_\psi > 3 m_{Z'}$. Similar to the previous decay, imposing $\tau_\psi > 20 \tau_U$, we get
\begin{equation}\label{eq:lifetime-bound2}
    \theta_{\nu\psi}^2 < 1.2 \times 10^{-30} \left(\frac{m_{Z'}}{10 {\rm~keV}}\right)^2 \left(\frac{10^{-4}}{g}\right)^2 \left(\frac{40 {\rm~keV}}{m_\psi}\right)^3\,,
\end{equation}
which is much stronger than the one in eq.~\eqref{eq:lifetime-bound}. This implies that the regime $m_\psi > m_{Z'}$ can work only with an effectively vanishing value of ${m_D}'$.

Sterile neutrino DM that mixes with active neutrinos can decay sub-dominantly also via radiative decay at 1-loop level as $\psi \to \nu \gamma$. 
For DM with mass $m_\psi \sim \mathcal{O}(10)$~keV, this process leads to an observable monochromatic line in the X-ray frequency band. 
In the past two decades, a number of telescopes have scrutinized the sky in search of a signature of this kind (see e.g., sec.~5.4 of ref. \cite{Abazajian:2021zui})
and ruled out part of the parameter space of sterile neutrinos, setting a constraint in the ($m_\psi$, $\theta_{\nu \psi}$) plane.\footnote{In 2014, two different groups claimed to have found a signal of this kind in different datasets \cite{Bulbul:2014sua,Boyarsky:2014jta,Boyarsky:2014ska}. However, to date, not only the interpretation of the claimed signal as being produced from decay of sterile neutrino DM, but even its effective presence in the data is still a matter of debate in the community \cite{Dessert:2023fen}.}
The observable measured in this case is the flux of photons detected by the telescope pointing towards DM-dominated objects. The flux of emitted photons is proportional to the decay width \cite{Pal:1981rm}
\begin{equation}
    \Gamma_{\psi \to \nu \gamma} = \frac{9\, \alpha\, G_F^2}{256 \cdot 4 \pi^4} \sin^2(2 \theta_{\nu \psi}) \, m^5_\psi \,.
    \label{eq:x-raywidth}
\end{equation}
Using the data put together in ref.~\cite{An:2023mkf}, one can estimate the constraint on $m_\psi$ and $\theta_{\nu \psi}$ to be 
\begin{equation}
    \theta^2_{\nu \psi} \lesssim 7.65 \times 10^{-13} \left( \frac{15 \text{ keV}}{m_\psi} \right)^5\,.
    \label{eq:x-raybound}
\end{equation}
Hence, X-ray limits can be satisfied by choosing $\theta_{\nu\psi}$ sufficiently small. Comparing with eq.~\eqref{eq:lifetime-bound}, unless $v_\phi$ is much larger than 2~GeV, we see that if the lifetime bound is satisfied, the X-ray bound \eqref{eq:x-raybound} is also satisfied. Similarly, if the lifetime bound of eq.~\eqref{eq:lifetime-bound2} is satisfied, the X-ray bound is always satisfied. 

Notice that in our framework, the production of DM happens independently of its mixing with active neutrinos. Therefore, $\theta_{\nu \psi}$ can always be chosen small enough to satisfy the strong upper bounds of eqs.~(\ref{eq:lifetime-bound}, \ref{eq:x-raybound}) without affecting the DM abundance. In other words, this implies that we can safely neglect any contribution to the DM abundance from its production from active neutrinos via the Dodelson-Widrow mechanism.

\subsection{Bounds on DM from structure formation}
\label{sec:structure-formation}

In our model we find typical dark matter masses in range of 10 to 100~keV, and therefore our candidate is subject to constraints from structure formation, as any other warm DM (WDM) candidate. Potentially large free-streaming scales prevent the formation of structure at small scales, leading to a suppressed matter power spectrum at small scales. The impact of WDM particles on structure formation depends essentially on two factors: the mechanism through which they are produced that determines their momentum distribution, and their mass. Typically, for each production mechanism, one can extract a lower bound on the mass of the DM candidate from different large-scale structure observables.

The impact of WDM candidates on structure formation is conventionally parametrized by considering the half-mode halo mass $M_{\rm hm}$ or half-mode length-scale $\lambda_{\rm hm}$ at which the power spectrum becomes suppressed by a factor 2 compared to the reference cold dark matter case. They are related by 
\begin{align}
    M_{\rm hm} &= \frac{4\pi}{3}\rho_{\rm DM}\left(\frac{\lambda_{\rm hm}}{2}\right)^3
     \approx 1.9\times 10^7\,M_\odot \left(\frac{\lambda_{\rm hm}}{0.1\, \rm Mpc}\right)^3
\end{align}
where $\rho_{\rm DM}$ is the dark matter energy density today. The limit on 
$M_{\rm hm}$ depends on the used cosmological data and analysis assumptions. For instance,
ref.~\cite{Zelko:2022tgf} obtains upper bounds at 95\%~CL in the range ${\rm log}_{10}(M_{\rm hm}/M_\odot)=7.0-8.6$, whereas ref.~\cite{Enzi:2020ieg} finds $M_{\rm hm} < 4.3\times 10^7M_\odot$. In the following we will adopt $\lambda_{\rm hm} < 0.1$~Mpc as a rough estimate for consistency with structure formation.

In order to estimate the suppression scale $\lambda_{\rm cutoff}$ in our model we follow closely the discussion in \cite{Berlin:2018ztp}. Our DM candidate comes into chemical equilibrium with the thermal bath of SM neutrinos, $Z'$ and massless $\chi$ and therefore will acquire a thermal momentum distribution. After the chemical decoupling, its co-moving abundance is constant, but it remains in thermal contact with the dark radiation via elastic $\psi\chi\leftrightarrow\psi\chi$ interactions. The cutoff scale $\lambda_{\rm cutoff}$ is determined when the DM decouples also kinetically from the dark radiation, see e.g.~\cite{Bringmann:2016ilk}. Let us denote the photon temperature at kinetic DM decoupling by $T_{\rm kd}$. Then there are two physically different effects which set $\lambda_{\rm cutoff}$: $(i)$ After kinetic decoupling, the DM will free-stream and suppress the logarithmic growth of structure before matter-radiation equality at co-moving scales smaller than
\begin{align}\label{eq:l_FS}
    \lambda_{\rm FS} &\approx \frac{1}{2} \int_{t_{\rm kd}}^{t_{\rm MRE}} dt\frac{v_\psi}{a(t)} 
    \approx \frac{1}{2}\left(\frac{4\pi^3g_{\rm eff}}{135}\right)^{-1/2} \sqrt{\frac{\xi}{T_{\rm kd}m_\psi}}\frac{M_{\rm pl}}{T_0} \log\frac{T_{\rm kd}}{T_{\rm MRE}} \,.
\end{align}
Here MRE refers to matter-radiation equality, $v_{\psi}$ is the DM velocity, $a(t)$ is the cosmic scale factor, $\xi$ is the ratio of the dark sector and photon temperatures, and $T_0=2.7$~K is the photon temperature today. The last approximation assumes that the DM particle is non-relativistic at kinetic decoupling \cite{Berlin:2018ztp}, such that $v_\psi = p_\psi/E_\psi \approx \sqrt{3 \xi T_{\rm kd}/m_\psi}(T/T_{\rm kd})$ for $T<T_{\rm kd}$.
$(ii)$~As long as the DM is kinetically coupled to the dark radiation bath, modes with wavelength smaller than the horizon will undergo acoustic oscillations \cite{Cyr-Racine:2013fsa}, preventing the formation of structure at these scales. This effect introduces another cutoff scale set by 
\begin{equation}\label{eq:l_AO}
    \lambda_{\rm AO} = \int_0^{t_{\rm kd}} \frac{dt}{a(t)} = \left. \frac{1}{aH}\right|_{\rm kd} 
    \approx \left(\frac{4\pi^3g_{\rm eff}}{45}\right)^{-1/2} \frac{M_{\rm pl}}{T_{\rm kd}T_0} \,.
\end{equation}
As a rough requirement for the consistency of our DM candidate with structure formation we then require 
\begin{equation}\label{eq:l_cutoff}
    \lambda_{\rm cutoff} = {\rm max}(\lambda_{\rm FS},\lambda_{\rm AO}) < 0.1\,\rm Mpc \,.
\end{equation}

\begin{figure}[t]
    \centering
    \includegraphics[width=0.49\textwidth]{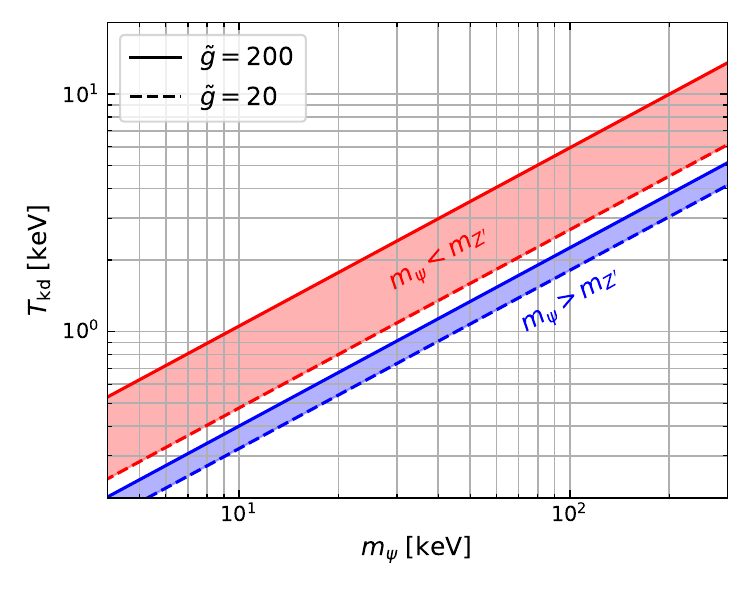}
    \includegraphics[width=0.49\textwidth]{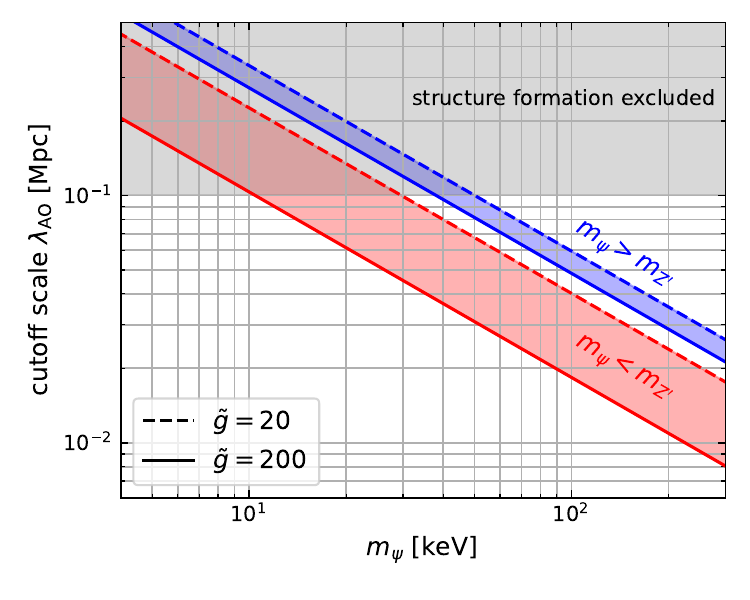}
    \caption{\textit{Left:} Photon temperature at kinetic decoupling of DM from the dark radiation bath. \textit{Right:} Cutoff scale due to acoustic oscillations $\lambda_{\rm AO}$. In both panels we used the requirement of correct DM abundance and determined $v_\phi$ via eq.~\eqref{eq:v_phi} either for $m_{Z'} \ll m_\psi$ (blue) or $m_{Z'} \gg m_\psi$ (red). The shaded bands indicate the range between two values of the massless degrees of freedom $\tilde g = 4N_\chi+2$, where the dashed (solid) boundary of the bands correspond to $\tilde g = 20\,(200)$. }
    \label{fig:Tkd}
\end{figure}

Kinetic decoupling has been studied for a range of simplified models in ref.~\cite{Bringmann:2016ilk}. In order to estimate $T_{\rm kd}$ we adopt the approximate expressions developed in the appendix of \cite{Bringmann:2016ilk}. In our model, $T_{\rm kd}$ depends on the VEV $v_\phi$, the DM mass $m_\psi$, and the massless degrees of freedom $\tilde g$, where the latter determines also the ratio of the dark sector-to-photon temperatures $\xi$, see eq.~\eqref{eq:xi} below.  Requiring the correct DM freeze-out abundance fixes $v_\phi$ for given $m_\psi$ and $\tilde g$, see eq.~\eqref{eq:v_phi}. We show $T_{\rm kd}$ as a function of $m_\psi$ and two representative values of $\tilde g$ in the left panel of fig.~\ref{fig:Tkd}. For DM masses in the range 10 to 100~keV we find decoupling temperatures 0.3~keV~$\lesssim T_{\rm kd} \lesssim 10$~keV, depending also on the hierarchy between $m_{Z'}$ and $m_\psi$. With this result we can calculate the free-streaming and acoustic oscillation cutoff scales from eqs.~\eqref{eq:l_FS} and \eqref{eq:l_AO}. For the relevant parameter regions we find that $\lambda_{\rm AO}\gtrsim\lambda_{\rm FS}$ and therefore $\lambda_{\rm AO}$ determines the cutoff scale. From the rough criterion in eq.~\eqref{eq:l_cutoff} we find from the right panel of fig.~\ref{fig:Tkd} that the lower bound on the DM mass in our model is somewhere in the range from 10~keV to 50~keV, depending on the $Z'$-$\psi$ mass hierarchy and the number of massless degrees of freedom. For increasing $\tilde g$ the lower bound decreases, as the kinetic decoupling occurs at higher temperatures. As we will see below, $\tilde g\approx 20$ is in tension with current limits on $N_{\rm eff}$ and therefore we require larger values than this. For $\tilde g=200$ (corresponding to $N_\chi \approx 50$) we find that our model is consistent with structure formation for DM masses $m_\psi \gtrsim 10\,(40)$~keV for $m_\psi < m_Z$ ($m_\psi > m_Z$).

\section{Relaxing the neutrino mass bound and predictions for $N_{\rm eff}$}
\label{sec:neutrino-mass}

Let us now combine this DM production mechanism with the scenario of ref.~\cite{Escudero:2022gez} to relax the neutrino mass bound from cosmology. In sec.~\ref{sec:ps} we review the relevant constraints on the parameter space of the model, whereas in sec.~\ref{sec:neff} we discuss the thermalization of the dark sector, the relaxation of the neutrino mass bound and our predictions for $N_{\rm eff}$.

\subsection{Constraints and viable parameter space}\label{sec:ps}

The interactions between $Z'$ and SM neutrinos are essentially unchanged from those of ref.~\cite{Escudero:2022gez}, as the mixing between neutrinos and $\psi$ is very suppressed. Therefore, the constraints on $Z' \r \nu\nu$ to determine the viable parameter space are also unchanged. In the following, for completeness, we review the most relevant constraints for the scenario.\\

\noindent\textbf{Thermalization:} From the discussion above, neutrinos should thermalize with the $Z'$ boson in the temperature interval 0.7 MeV $> T_\gamma >$ 10 eV. The thermalization constraint is applied to the interaction $Z' \r \nu{\nu}$ 
rather than $Z' \r \nu\chi$, even though the former is suppressed by $\theta_{\nu\chi}$ as initially there are no $\chi$ present in the plasma. Since the thermally averaged decay rate peaks at $T \sim {m_{Z'}}/{3}$, we obtain a lower bound on the coupling $\lambda_{Z'}^{\nu \nu}$ by requiring the interaction rate $\braket{\Gamma(Z' \r \nu\nu)} > H(T\sim m_{Z'}/3)$.
Using the expression for the Hubble rate in the radiation dominated universe, we obtain
\begin{equation}\label{eq:thermalbound}
   \lambda_{Z'}^{\nu \nu} \gtrsim 3.1 \times 10^{-12} \left({\frac{m_{Z'}}{\text{keV}}}\right)^{1/2} \,.
\end{equation}
The region where neutrinos do not thermalize with $Z'$ in the early universe is highlighted in light red in fig.~\ref{fig:ps}.\\

\noindent\textbf{BBN constraints:} In order not to spoil BBN, $Z'$ must not be in thermal equilibrium with neutrinos at $T > 0.7$ MeV, otherwise it would deplete the population of neutrinos which participate in $p \leftrightarrow n$ conversions and would also contribute to the expansion rate of the universe along with the $\chi$ particles. Therefore, we require $\langle \Gamma (Z' \leftrightarrow \nu \nu) \rangle < H(T = 0.7 \text{ MeV})$, which leads to
\begin{equation}
    \lambda_{Z'}^{\nu \nu} \lesssim 10^{-7} \left(\frac{\text{keV}}{m_{Z'}}\right) \,. 
\end{equation}
This constraint is shown as the light blue region in fig.~\ref{fig:ps}. Further, if a minimal abundance of $\chi$ is present in the plasma before BBN (for example, having been produced from decays of $N$ or $N'$), in order to avoid its exponential growth before BBN we need to impose 
\begin{equation}
    \langle \Gamma ( \nu \chi \leftrightarrow \chi \chi) \rangle < H(T = 0.7 \text{ MeV}) \,,
\end{equation}
which gives a bound on the coupling
\begin{equation}
    g \lesssim  10^{-3} \left(\frac{\theta_{\nu \chi}}{10^{-3}}\right)^{-1/2}\,, 
\end{equation}
excluding the dark blue shaded region in fig.~\ref{fig:ps}.\\

\noindent\textbf{CMB constraints:} In order for the CMB not to be distorted by $\nu \nu \to Z'$ and $ Z' \to \chi \chi$ interactions, they must be rendered inefficient at $z < 10^5$, to be free of CMB constraints coming from neutrino free-streaming and power spectra \cite{Taule:2022jrz}, therefore, we require\footnote{We use the standard relation between temperature and redshift, $T(z) = T_0 (1+z)$, where $T_0 = 2.3 \times 10^{-4}$ eV is today's CMB temperature, and redshift $z = 10^5$ corresponds to $T(10^5) = 23$~eV.}
\begin{align}
    &\langle \Gamma (\nu \nu \leftrightarrow Z') \rangle < H(T = 23 \text{ eV}) \,.
\end{align}
Additionally, as $Z'$ decays to a lot of massless states which are relativistic, we also require
\begin{align}
    \langle \Gamma (Z' \leftrightarrow \chi + \chi / \nu) \rangle < H(T = 23 \text{ eV})\,,
\end{align}
which excludes the region shaded in light green in fig.~\ref{fig:ps}.
Finally, to ensure that CMB is not perturbed by the lack of free-streaming of $\chi$, analogous to the free-streaming of $\nu$, we impose 
\begin{equation}
    \langle \Gamma ( \chi \chi \leftrightarrow \chi \chi) \rangle < H(T = 23 {\rm~eV}) \,,
\end{equation}
which gives us a constraint: $v_\phi \gtrsim 2 \times 10^{4}$ keV, shown by the bright green region in fig.~\ref{fig:ps}.\\

\noindent\textbf{Constraints on active-sterile neutrino mixing:} The $\nu-\chi$ mixing would also lead to the production of $\chi$ via oscillations in the early universe, well before the neutrinos decouple and BBN, therefore, the $\chi$ will contribute to the energy density of the universe and change $\neff$. Therefore, requiring $\Delta \neff < 0.3$, a bound can be placed on the mixing angle~\cite{Escudero:2022gez}
\begin{align}
    \abs{\theta_{\nu\chi}} \leq 10^{-3}\left({\frac{10}{N_\chi}}\right)^{1/2}\left({\frac{0.2{\rm~eV}}{m_\nu}}\right)^{1/2}\,.
\end{align}
However, recently it was shown that considering the $\chi$ self-interactions \cite{cite-key}, it is possible to have larger mixing angles, leading to the possibility of detecting $\chi$ from neutrino oscillation experiments. Varying the parameters of interest, one expects $\theta_{\nu\chi} \leq 10^{-4}-{10^{-1}}$. The constraints from $\nu-\chi$ after considering the $\chi$ self-interactions are shown by the gray region in the top plot of fig.~\ref{fig:ps}, and is applicable only for $\theta_{\nu\chi}>10^{-3}$. We find that the parameter space corresponding to the extreme limits, i.e., $\theta_{\nu\chi} = 10^{-1},10^{-4}$ is quite constrained and incompatible with successfully accommodating a DM candidate, hence we choose $\theta_{\nu\chi} = 10^{-3},10^{-2}$ in fig.~\ref{fig:ps}.

\subsection{Suppression of neutrino mass, $N_\text{eff}$ and dark sector temperature}\label{sec:neff}

Let us discuss now the equilibration of the dark sector in somewhat more detail. We assume that the new degrees of freedom come in equilibrium instantaneously with the active neutrinos when they have a temperature $T_\nu^{\rm eq}$ and form a thermal system with a new temperature $T_{\rm eq}$, with $m_{Z'},m_\psi \ll T_{\rm eq} \ll 1$~MeV.\footnote{Note that our masses are $\mathcal{O}(10-100)$~keV and actually the hierarchy is mild. Nevertheless, we take $\psi$ and $Z'$ as fully relativistic at equilibration for the estimates in this section. We have checked that they are in good agreement with the solution of the Boltzmann equations.} For the parameter region of interest, the following interactions come into thermal equilibrium:
\begin{align}
     &Z' \leftrightarrow ff \,, \quad Z' \leftrightarrow ff' \,, 
     \label{eq:Zreact-12}\\ 
     & Z'Z' \leftrightarrow f f \,, 
     \label{eq:Zreact-22}\\ 
     &ff \leftrightarrow f'f' \,,
     \label{eq:freact-22}
\end{align}
with $f, f' = \nu, \chi, \psi$. As discussed above, the reaction $Z'\leftrightarrow \nu\nu$ is crucial to equilibrate the dark sector at first place, however, the less suppressed reactions with $f = \chi,\psi$ will be much faster once these particles are abundant. In particular, both, $1\leftrightarrow 2$ and $2\leftrightarrow 2$ processes involving $Z'$ will simultaneously come into equilibrium. This implies that all chemical potentials are zero and there is no conserved particle number.\footnote{This is different from the scenario considered in refs.~\cite{Escudero:2022gez,Farzan:2015pca}, where only the $1\leftrightarrow2$ prozesses from eq.~\eqref{eq:Zreact-12} have been taken into account, but not the $2\leftrightarrow2$ prozesses from eq.~\eqref{eq:Zreact-22}. In that case, there is a conserved particle number 
(fermions $f$ are created or destroyed only in pairs, while $Z'$ numbers change by one unit) and chemical potentials develop, see also the discussion in Appendix~B of \cite{Escudero:2022gez}. In contrast, in our model, the processes \eqref{eq:Zreact-22} proceed through $t$- and $u$-channel diagrams with $\sigma \sim g^4/T^2$, for which the interaction rates for coupling strength of $10^{-5} \lesssim g \lesssim 10^{-3}$ (c.f.\ fig.~\ref{fig:ps}) are much larger than the expansion rate in the relevant temperature range. \label{ft:cons}} 

Hence the system is fully characterized by its temperature $T_{\rm eq}$, which is obtained from energy conservation during equilibration:
\begin{align}
    \rho_\nu(T_\nu^{\rm eq}) = 
    \sum_{f=\nu,\chi,\psi} \rho_f(T_{\rm eq}) + 
    \rho_{Z'}(T_{\rm eq}) \,, \label{eq:e-cons-eq}
\end{align}
with the energy density of a species $i$ given by 
\begin{align}
    \rho_i(T) = \left[ \frac{7}{8}\right] \frac{\pi^2}{30} g_i T^4 \,, \label{eq:rho}
\end{align}
where the factor in square-brackets holds for fermions and should be dropped for bosons. Eq.~\eqref{eq:e-cons-eq} determines the ratio $T_{\rm eq}/T_\nu^{\rm eq}$ in terms of the involved degrees of freedom. Then the system evolves adiabatically to a temperature $T_{\rm fin} \ll m_\psi, m_{Z'}$, when the DM $\psi_L$ and $Z'$ particles have become non-relativistic and therefore no longer contribute to dark radiation. Here we can use conservation of co-moving entropy:
\begin{align}
    a_{\rm eq}^3 s_{\rm eq}(T_{\rm eq}) = 
    a_{\rm fin}^3 s_{\rm fin}(T_{\rm fin})  \,, 
    \label{eq:s-cons-fin}
\end{align}
with the cosmic scale factor $a$ and $s = 4\rho / (3T)$ for massless particles. At the right-hand side of eq.~\eqref{eq:s-cons-fin} we take into account that only neutrinos and the massless degrees of freedom ($\chi_L,\chi_R, \psi_R$) contribute at $T_{\rm fin}$. 
Combining eqs.~\eqref{eq:e-cons-eq} and \eqref{eq:s-cons-fin} leads to the ratio of the dark sector temperature $T_{\rm dark}$ to the neutrino temperature in the SM (evaluated at $T_{\rm dark} = T_{\rm fin}$) as
\begin{align} \label{eq:R}
    \frac{T_{\rm dark}}{T^{\rm SM}_\nu} = & 
    \left( \frac{g_\nu +  \tilde{g} + g_\psi + \frac{8}{7}g_{Z'}}
    {g_\nu + \tilde{g}} \right)^{1/3} \times \left(\frac{g_\nu}{g_\nu + \tilde{g} + g_\psi + \frac{8}{7}g_{Z'}}\right)^{1/4} 
\end{align}
and the dark sector-to-photon temperature as
\begin{equation}\label{eq:xi}
    \xi \equiv \frac{T_{\rm dark}}{T_\gamma} = \left(\frac{4}{11}\right)^{1/3}     \frac{T_{\rm dark}}{T^{\rm SM}_\nu} \,.
\end{equation}
Here $\tilde{g} = 4N_\chi +2$ are the massless degrees of freedom and $g_\psi = 2 , g_{Z'}=3$, $g_\nu=6$. Note that for the scalar in our model we expect masses of order of the dark VEV, with 10~MeV~$\lesssim v_\phi \lesssim 10$~GeV. Therefore, the scalar will not be present in the dark sector thermal bath in the relevant temperature range $T\lesssim 1$~MeV and does not contribute to the degrees of freedom here. For $N_{\rm eff}$ we then obtain
\begin{align}\label{eq:Neff-pred}
    N_{\rm eff} &\equiv  \frac{8}{7} \left(\frac{11}{4}\right)^{4/3} \frac{\rho_{\rm dark}}{\rho_\gamma}  
 = \frac{g_\nu + \tilde{g}}{2} \left(\frac{T_{\rm dark}}{T^{\rm SM}_\nu}\right)^4 \,. 
\end{align}
Using eq.~\eqref{eq:R}, we see that this expression for $N_{\rm eff}$ converges to $g_\nu/2 = 3$ for both, $\tilde{g},g_\psi,g_{Z'}\to 0$ as well as for $\tilde{g} \to \infty$. Note that this is the value for $N_{\rm eff}$ relevant for CMB observables, whereas for BBN $N_{\rm eff}$ in our scenario should be very close to the SM value. For the suppression of the neutrino number density we obtain
\begin{align}\label{eq:n-ratio}
    \frac{n_\nu}{n_\nu^{\rm SM}} &= \left(\frac{T_{\rm dark}}{T^{\rm SM}_\nu}\right)^3  = \frac{g_\nu + \tilde{g} + g_\psi + \frac{8}{7}g_{Z'}}
    {g_\nu + \tilde{g}} 
    \left(\frac{g_\nu}{g_\nu + \tilde{g} + g_\psi + \frac{8}{7}g_{Z'}}\right)^{3/4} \,.
\end{align}

\begin{figure}[!t]
    \centering
    \includegraphics[width=\textwidth]{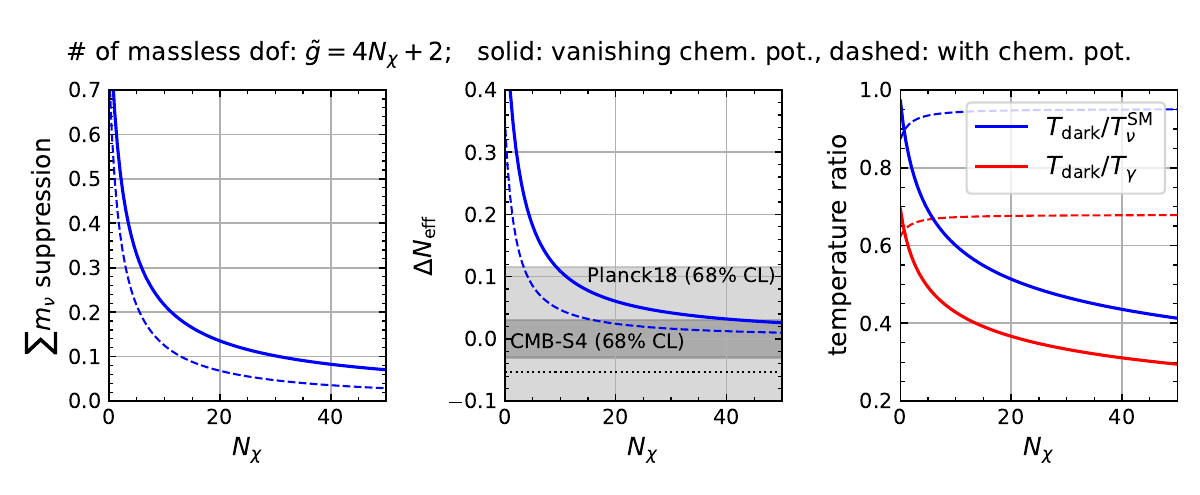}
    \caption{\textit{Left:} the ratio $n_\nu/n_\nu^{\rm SM}$, i.e., the effective suppression factor of $\sum m_\nu$, see eqs.~\eqref{eq:sum-suppression},\eqref{eq:n-ratio}, \textit{middle:} $\Delta\neff$, \textit{right:} ratios of the dark-sector temperature to the neutrino temperature in the SM and to the photon temperature as a function of the number of $\chi$ generations $N_\chi$, which is related to the number of massless degrees of freedom in our model by $\tilde{g} = 4N_\chi+2$. Dashed curves correspond to the case with particle number conservation, shown for illustration purposes. The dotted line in the middle panel indicates the current \emph{Planck} best fit point $\Delta\neff = 2.99 - 3.044=-0.054$ \cite{Planck:2018vyg} with the $1\sigma$ interval $\pm 0.17$ shown with light gray shading, and the dark shaded region indicates the sensitivity at $\pm 1\sigma$ of a future CMB-S4 mission \cite{Abazajian:2019eic}.}
    \label{fig:Neff}
\end{figure}
The results of this calculation are shown in fig.~\ref{fig:Neff}, where for comparison we show with dashed curves also the case when particle numbers are conserved, as considered in refs. \cite{Farzan:2015pca,Escudero:2022gez}, see footnote~\ref{ft:cons}. 
The latter are calculated following the procedure outlined in Appendix~B in ref. \cite{Escudero:2022gez}.
For given $N_\chi$ we observe a slightly smaller suppression of the neutrino mass bound and a larger contribution to $N_{\rm eff}$ than in the case with particle number conservation. For large $\tilde{g}$, the suppression factor in eq.~\eqref{eq:n-ratio} decreases like $\tilde{g}^{-3/4}$, compared to the 
factor $(1+ \tilde{g}/6)^{-1}$ for the case with number conservation \cite{Farzan:2015pca,Escudero:2022gez}. 

Our scenario predicts a sizable deviation of $N_{\rm eff}$ from the SM value at recombination. From the middle panel of fig.~\ref{fig:Neff} we see that small values of $N_\chi \lesssim 10$ are already in tension with the \emph{Planck} result $\neff = 2.99 \pm 0.17$ \cite{Planck:2018vyg}. For the range
$10\lesssim N_\chi \lesssim 40$ (where also relevant relaxation of the neutrino mass bound occurs), deviations of $\neff$ from the SM value may be observable by future CMB missions, which should reach sensitivities of 
0.07 \cite{SimonsObservatory:2018koc} or even
0.03 \cite{Abazajian:2019eic} at $1\sigma$. See also ref. \cite{Upadhye:2024ypg} for a recent study of such dark radiation scenarios in the context of observables in the non-linear regime. Note that we predict an additional contribution to $\neff$ at CMB, whereas its value relevant for BBN should be largely unaffected in our model. Hence,  different $\neff$ values inferred from BBN and CMB would be another potential signature of our model. Let us also note that the analysis in ref.~\cite{Allali:2024cji} shows that recent DESI BAO measurements \cite{DESI:2024mwx} leading to the tight bound on $\sum m_\nu$ from eq.~\eqref{eq:cosmobound} can be better accommodated by adding some amount of dark radiation (i.e., $\Delta N_{\rm eff} > 0$), as predicted in our scenario. 

Finally, we also comment on the rather different behavior of the dark sector temperature in the two cases. For conserved particle numbers, the ratio of the dark sector temperature to the neutrino temperature in the SM saturates around 0.95 for large $\tilde{g}$, whereas in the case with zero chemical potential, eq.~\eqref{eq:R}
gives $1/\tilde{g}^{1/4}$ dependence for large $\tilde{g}$, leading to a significant cooling of the dark sector in our model.

\section{Summary and discussion}\label{sec:conc}

We have presented a simple extension of the SM featuring a light dark sector with a $U(1)$ gauge symmetry, embedded in a seesaw model for neutrino masses. The coupling of the dark sector to the SM occurs via mass-mixing of the dark fermions with neutrinos. The fermion spectrum contains heavy sterile neutrinos responsible for the seesaw mechanism, a large number of massless states $\chi$, and a DM candidate $\psi$ with mass in the 10 to 100~keV range. Both, active neutrino masses as well as the DM mass are generated via the type-I seesaw mechanism. We have identified regions of parameter space, where the following phenomenology is valid:
\begin{itemize}
    \item The light dark sector particles play no role in the universe for temperatures above 1~MeV but come into thermal equilibrium with the SM neutrinos \emph{after} neutrino decoupling from the SM but \emph{before} recombination. This sets the typical scale for the dark gauge boson mass $m_{Z'} \sim 10$~keV.
    \item At late times the number density of active neutrinos is suppressed compared to the standard evolution, which allows to accommodate sizable neutrino masses with cosmology. The amount of neutrino suppression is determined by the number of massless degrees of freedom in the dark sector, which act as dark radiation during recombination.  
    \item The DM fermion $\psi$ first thermalizes together with the light dark sector, but then freezes out when the dark-sector temperature drops below its mass. Hence, the DM relic abundance is set by the dark sector gauge interactions and is independent of the DM mixing with SM neutrinos, which can be taken to be very small in order to satisfy constraints on DM stability, avoiding also measurable X-ray signatures.  
\end{itemize}

\begin{table}[!t]
    \centering
    \resizebox{\linewidth}{!}{
    \begin{tabular}{ccccc|cccccc|cc}
        \hline \hline
           $M{\rm~[GeV]}$ & $M'{\rm~[GeV]}$ & $m_D{\rm~[GeV]}$ & $\kappa'{\rm~[GeV]}$ & $\Lambda{\rm~[GeV]}$ & $v_\phi{\rm~[GeV]}$ & $m_{\psi}{\rm~[keV]}$ & $m_{Z'}{\rm~[keV]}$ & $g = m_{Z'}/v_\phi$ & $\theta_{\nu\chi}$ & $N_\chi$ & $n_\nu/n_\nu^{\rm SM}$ & $\Delta \neff$\\
          \hline
          $10^{11}$ & $10^2$ & 4.47 & 0.043 & 0.004 & 0.5 & 18.5 & 100 & $2\times 10^{-4}$ & $10^{-3}$ & 10 & 0.216 & 0.109\\
          $10^{12}$ & $10^3$ & 14.14 & 0.23 & 0.141 & 0.8 & 53 & 77 & $9.6\times 10^{-5}$ & $10^{-2}$ & 10 & 0.216 & 0.109\\
          $10^{13}$ & $10^2$ & 44.7 & 0.1 & 0.044 & 0.6 & 100 & 32 & $5.3\times 10^{-5}$ & $10^{-3}$ & 20 & 0.135 & 0.060\\
          \hline \hline
    \end{tabular}}
    \caption{A set of benchmark points for the model where the mechanism to reduce the cosmological neutrino number density works and the observed DM relic abundance ($\Omega_\psi h^2 = 0.12 \pm 0.0012$) is obtained. The left columns show the relevant quantities from the fermion mass matrix and illustrate that our assumed hierarchy eq.~\eqref{eq:hierarchy} is satisfied. The middle columns indicate our main parameters of interest (see eq.~\eqref{eq:param}), and the right columns correspond to the effective suppression factor of $\sum m_\nu$ for cosmology and our predictions for $\Delta\neff$, respectively. For all three points the active neutrino mass $m_\nu = m_D^2/M$ is 0.2~eV.}
    \label{tab:bps}
\end{table}

Although the particle content of the model is complex, its free parameters are not numerous. Indeed, if we assume that the value of mixing between $\nu$ and $\chi$ is the same for all $N_\chi$ $\chi$ species, the phenomenology of the model is essentially determined by five independent parameters (see eq.~\eqref{eq:param}), which makes the model rather predictive.
The parameter space of the model is well constrained and closed from all sides in many directions, see e.g., fig.~\ref{fig:ps}. In Tab.~\ref{tab:bps} we give three benchmark points where all the above requirements are met. The left part of the table shows the relevant entries of the fermion mass matrix, which satisfy the assumed hierarchies from eqs.~\eqref{eq:hierarchy} and \eqref{eq:hier2}. The parameters $m'_D,\Lambda',\kappa$ can be taken arbitrarily small. In particular, a very small value of $m'_D$ is required in order to have tiny mixing of the DM fermion with active neutrinos, as $\theta_{\nu\psi}= m'_D/\kappa'$. This is required in order to suppress the decays $\psi\to \nu\chi\chi$ and $\psi\to Z'\nu$ (if kinematically allowed) to have a sufficiently long-lived DM particle. 

In the middle part of the table we provide the values of the VEV breaking the dark U(1) symmetry $v_\phi$. Noting that the entries $\kappa'$ and $\Lambda$ are related to this VEV by Yukawa couplings (see eq.~\eqref{eq:yukawas}), we confirm that these couplings remain within the perturbative regime. Typical gauge couplings are in the range of $10^{-5}$ to $10^{-4}$. The mixing of the active neutrinos with the massless states $\theta_{\nu\chi}=\Lambda/m_D$ is constrained roughly as 
$10^{-3} \lesssim \theta_{\nu\chi}\lesssim 10^{-2}$, potentially testable in oscillation experiments only by saturating the upper bound.  

The number of massless degrees of freedom in the dark sector, $\tilde g$ is related to the number of generations of $\chi$ fermions by $\tilde g = 4N_\chi + 2$ and sets the suppression factor of the neutrino density via eq.~\eqref{eq:n-ratio}, which corresponds to the factor about which a bound on the neutrino mass from cosmology is relaxed. For instance, for the three benchmark points in table~\ref{tab:bps}, the DESI bound $\sum m_\nu < 0.071$~eV from eq.~\eqref{eq:cosmobound} would be relaxed to 0.33, 0.33, 0.52~eV, respectively, making it safely compatible with the lower bounds from the neutrino oscillation data for both mass orderings and consistent with the  value $m_\nu = 0.2$~eV assumed for the benchmark points.

While our specific DM candidate is very difficult to test observationally, the over-all model leads to indirect predictions. In particular, in our scenario we expect an increased energy density of dark radiation, expressed in terms of the effective number of neutrino generations $N_{\rm eff}$. This effect emerges due to the disappearance of the massive $Z'$ and DM particles from the relativistic bath, which leads to an effective heating of the dark sector. The amount of increase in $N_{\rm eff}$ decreases with $\tilde g$ (see eq.~\eqref{eq:Neff-pred} and fig.~\ref{fig:Neff}). Values of $N_\chi \lesssim 10$ are already in tension with the present CMB constraint on $N_{\rm eff}$, whereas $N_\chi$ up to about 40 may lead to sizeable values in reach of planned CMB missions such as the Simons Observatory~\cite{SimonsObservatory:2018koc} or CMB-S4~\cite{Abazajian:2019eic}. With DM masses in the 10 to 100~keV range, we predict a suppression of the matter power spectrum at small scales. For instance, the three benchmark points from table~\ref{tab:bps} lead to suppression of structure formation at scales below 0.11, 0.051, and 0.054~Mpc, respectively, with the first point marginally consistent with observations, see sec.~\ref{sec:structure-formation} for the corresponding discussion. 

To conclude, the comparison of cosmological and laboratory determinations of neutrino masses may reveal exciting signs of new physics, if near future cosmological observations continue present tendencies and lead to increasingly strong bounds on the sum of neutrino masses and/or upcoming laboratory measurements of the absolute neutrino mass establish a non-zero value. We have presented an economical UV complete model offering an explanation in such a case. In addition to making large neutrino masses consistent with cosmology our model provides a simple seesaw explanation of neutrino masses, it incorporates a dark matter candidate in the 10 to 100~keV mass range in a straightforward way in the dark-fermion spectrum and its abundance is obtained naturally by freeze-out in the dark sector.
Genuine predictions of the model is an increased value of the relativistic degrees of freedom at late times, potentially observable by future CMB observations and a suppressed matter power spectrum at small scales, as characteristic for warm DM candidates.

\acknowledgments We would like to thank Miguel Escudero and Jorge Terol Calvo for many useful discussions. DV also thanks Juan Herrero for his help. CB would like to thank Evgeny Akhmedov for his precious clarifications about sterile neutrinos, and Ioana Zelko for the fruitful discussion regarding limits on WDM from structure formation and for her availability to help. TS thanks Toshi Ota for useful discussions. DV is supported by the ``Generalitat Valenciana” through the GenT Excellence
Program (CIDEGENT/2020/020) and CIBEFP/2022/31, and greatly acknowledges the hospitality of Theoretical Astroparticle Physics group (IAP) at KIT, where the initial part of the project was carried out.
This work has been supported by the European Union’s Framework Programme for Research and Innovation Horizon 2020 under grant H2020-MSCA-ITN-2019/860881-HIDDeN.  

\appendix

\section{Diagonalisation of mass matrix}\label{sec:diag}

The matrix in eq.~\eqref{eq:massmatrix} can be reduced to a $2 \times 2$ form and block-diagonalised with a unitary transformation,
\begin{align}
    {\mathbf A}^T \begin{pmatrix}
        0 & \Upsilon \\
        \Upsilon^T & \mathbf{M}
    \end{pmatrix}{\mathbf A} = \begin{pmatrix}
        \hat{\mathbf{m}} & 0\\
        0 & \hat{\mathbf{M}}
    \end{pmatrix}\,,
\end{align}
where $\Upsilon$ is the $3 \times 2$ matrix containing the couplings and $\mathbf{M}$ is the diagonal matrix containing heavy neutrino masses. We choose the following Ansatz for $\mathbf{A}$ \cite{Grimus:2000vj}
\begin{equation}
    \mathbf{A} = \begin{pmatrix}
        \mathds{1}_{3 \times 3} & {\rho} \\
        {-\rho^\dagger} & \mathds{1}_{2\times 2}
    \end{pmatrix}\,,
\end{equation}
with the small mixing parameter $\rho$ given by $\Upsilon^\ast (M^{-1})^\dagger$, and
\begin{equation}
    \hat{\mathbf{m}} = -\Upsilon \mathbf{M}^{-1} \Upsilon^T\,,\quad \hat{\mathbf{M}} \sim \mathbf{M}\,.
\end{equation}
The resultant $3 \times 3$ matrix $\hat{\mathbf{m}}$ is a rank 2 matrix that can be further diagonalised to obtain the mass eigenvalues for the light fields
\begin{equation}
    \mathbf{R}^T \hat{\mathbf{m}}\, \mathbf{R} = {\rm diag}(m_\chi,m_{\nu},m_\psi)\,,
\end{equation}
where the rotation matrix $\mathbf{R}$ can be parameterized in terms of two mixing parameters ($\alpha,\beta$) as follows
\begin{align}
    \mathbf{R} = \begin{pmatrix}
        1 & \beta & 0\\
        -\beta^\dagger & 1 & \alpha\\
        0 & -\alpha^\dagger & 1
    \end{pmatrix}\,,
\end{align}
with $\beta \sim \Lambda^\ast/m_D^\ast$ and $\alpha \sim {m_D}'^\ast/\kappa'^\ast$. The mixing matrix $\mathbf{W}$ can then be expressed as a product of the matrices $\mathbf{A},\mathbf{R}$, and further $U_{\nu}$.
\section{Interaction rates and Boltzmann equations}\label{sec:rates}

The relevant expressions for $N$ and $Z'$ decay widths are
\begin{align}
    \Gamma_{N' \r lH/{\chi\phi}/{\psi\phi}} = \frac{M'}{8\pi}{Y'}_{\nu/\chi/\psi}^2\,,\qquad\Gamma_i \equiv \Gamma_{Z' \r ii} = N_i\,\frac{\lambda_{ii}^2}{24\pi} m_{Z'} \left(1-\frac{4m_i^2}{m_{Z'}^2}\right)^{3/2}\,,
\end{align}
where $N_i$ is the number of generations of $i$ Majorana fermions. The thermally averaged decay rates are given by \cite{EscuderoAbenza:2020cmq}
\begin{align}\label{eq:dmdecay}
    &\bra\Gamma_{a \r 1 +2}\ket = \Gamma_a \frac{K_1(m_a/T)}{K_2(m_a/T)}\,,\quad\bra\Gamma_{1+2 \r a}\ket = \Gamma_a \frac{g_a}{2g_1}(m_a/T)^2 K_1(m_a/T)\,,
\end{align}
where $K_{1,2}$ are the modified Bessel functions of the first and second kind and $g_{a,1}$ are the internal degrees of freedom of the decaying and daughter particle. 

The $2 \leftrightarrow 2$ cross-sections involving light/massless particles mediated via $Z'$ are 
\begin{align}
    &\sigma \sim \frac{g^4}{16\pi^2}\frac{1}{m_{Z'}^4}T^2\quad (m_{Z'}\gtrsim T)\,,\quad\sigma \sim \frac{g^4}{16\pi^2}\frac{1}{T^2}\quad (m_{Z'}\ll T)\,.
\end{align}
The rate and thermal average for $\psi$ to $\chi$ annihilation are given by
\begin{align}
    \sigma_{\psi\chi}(s) &= \tilde{N}\frac{g^4}{12\pi}\frac{(s+2m_\psi^2)(1-4m_\psi^2/s)^{-1/2}}{(s-m_{Z'}^2)^2+m_{Z'}^2 \Gamma_{Z'}^2}\,,\nonumber\\
    \braket{\sigma v}_{\psi\chi} &= \frac{1}{8T\,m_\psi^4 K_2\left(\frac{m_\psi}{T}\right)^2}\int_{s_{\rm min}}^\infty ds \, s^{3/2} K_1\left(\frac{\sqrt{s}}{T}\right) \lambda(1,m_\psi^2/s,m_\psi^2/s)\,\sigma_{\psi\chi}(s)\,,
\end{align}
where $s$ is the centre of mass energy squared, $\Gamma_{Z'}$ is the total decay width of the mediator $Z'$ and $\lambda(x,y,z)=x^2+y^2+z^2-2xy-2xz-2yz$ is the Källen function. In the region far from resonances, we can expand $\sigma v$ in power of $v$ using $s= 4m_\psi^2 + m_\psi^2 v^2$, and then take the thermal average to obtain eq.~\eqref{eq:dmann}.

Finally, the DM decay widths are given by
\begin{align}
    \Gamma_{\psi\r \nu\chi\chi} &\simeq \tilde{N}\,\frac{m_\psi^5}{96\pi^3 M_{Z'}^4}(\lambda^{\chi\chi}_{Z'})^2 (\lambda^{\nu\psi}_{Z'})^2\,,\nonumber\\
    \Gamma_{\psi \r Z' \nu} &\simeq \frac{g^2 \theta_{\nu\psi}^2\,m_\psi}{32\pi}\left(1-\frac{m_{Z'}^2}{m_\psi^2}\right)\left(1+\frac{m_\psi^2}{m_{Z'}^2}-\frac{2m_{Z'}^2}{m_\psi^2}\right)\,.
\end{align}

We work with the following set of coupled Boltzmann equations for the evolution of number densities of the various BSM fields 
   {\small \begin{align}\label{eq:be}
    \frac{dY_{\nu}}{dx}&=\frac{\braket{\Gamma_\nu}}{H x}\left(Y_{Z'}-Y_{Z'}^{\rm eq}\frac{Y_\nu^2}{{Y_\nu^{\rm eq}}^2}\right)\,,\nonumber\\
    \frac{dY_{Z'}}{dx}&=\sum_{i=\nu,\chi,\psi}-\frac{\braket{\Gamma_i}}{H x}\left(Y_{Z'}-Y_{Z'}^{\rm eq}\frac{Y_i^2}{{Y_i^{\rm eq}}^2}\right)+\frac{s \bra \sigma v \ket_{\psi\psi\to Z' Z'}}{H x}\,\left(Y_\psi^2 - \frac{{Y_\psi^{\rm eq}}^2}{{Y_{Z'}^{\rm eq}}^2}Y_{Z'}^2\right)\,,\nonumber\\
    \frac{dY_{\chi}}{dx}&=\frac{\braket{\Gamma_\chi}}{H x}\left(Y_{Z'}-Y_{Z'}^{\rm eq}\frac{Y_\chi^2}{{Y_\chi^{\rm eq}}^2}\right) + \frac{s\bra \sigma v \ket_{\psi\psi\to\chi\chi}}{Hx}\left(Y_\psi^2 - \frac{{Y_\psi^{\rm eq}}^2}{{Y_\chi^{\rm eq}}^2}Y_\chi^2\right)\,,\nonumber\\
    \frac{dY_{\psi}}{dx}&=\frac{\braket{\Gamma_\psi}}{H x}\left(Y_{Z'}-Y_{Z'}^{\rm eq}\frac{Y_\psi^2}{{Y_\psi^{\rm eq}}^2}\right) - \frac{s\bra \sigma v \ket_{\psi\psi\to\chi\chi}}{Hx}\left(Y_\psi^2 - \frac{{Y_\psi^{\rm eq}}^2}{{Y_\chi^{\rm eq}}^2}Y_\chi^2\right)-\frac{s \bra \sigma v \ket_{\psi\psi\to Z' Z'}}{H x}\,\left(Y_\psi^2 - \frac{{Y_\psi^{\rm eq}}^2}{{Y_{Z'}^{\rm eq}}^2}Y_{Z'}^2\right) \,,   
    \end{align}}%
where $Y_i \equiv n_i/s$ is the yield, $s = (2\pi^2/45)g_{\ast s}T_\gamma^3$ is the entropy density, $H(T_\gamma) = 1.66 \sqrt{g_{\rm eff}}(T_\gamma^2/M_{\rm pl})$ with $g_{\rm eff}=g_\gamma + g_{\rm dark}\xi^4$, and here we define $x \equiv m_{Z'}/T_\gamma$. The equilibrium yield of a particle with mass $m_i$ is given by
\begin{equation}
    Y_i^{\rm eq} = 0.115\,\frac{g_i}{g_{\ast s}}\,\xi\, x_i^2\, K_2(x_i/\xi)
\end{equation}
with $x_i = m_i/T_\gamma$. For massless particles $Y_\chi^{\rm eq} = 0.23(g/g_{\ast s})\,\xi^3$ and in the case of neutrinos corresponds to their yield after they have decoupled from the plasma. Note that the $2 \leftrightarrow 2$ processes involving $\nu$'s are quite suppressed by the small mixing angles, and we have checked that their contribution is negligible compared to the processes considered above in eqs.~\eqref{eq:be}. In the case $m_{Z'}<2m_\psi$, the terms corresponding to $\braket{\Gamma_\psi}$ are not taken into account.

We solve these equations by taking negligible abundances for all the dark sector particles as initial conditions. Further, it should be mentioned that we assume a fixed value of the dark sector temperature when solving the Boltzmann equations. A precise treatment should also involve the temperature evolution equations. We have checked that our results for the DM yield are not very sensitive to how we chose the dark sector temperature, and therefore we are confident that this is a reasonable approximation.

\bibliographystyle{JHEP}
\bibliography{refs}

\providecommand{\href}[2]{#2}\begingroup\raggedright\begin{thebibliography}{10}

\bibitem{Lesgourgues:2006nd}
J.~Lesgourgues and S.~Pastor, \emph{{Massive neutrinos and cosmology}},
  \href{https://doi.org/10.1016/j.physrep.2006.04.001}{\emph{Phys. Rept.}
  {\bfseries 429} (2006) 307}
  [\href{https://arxiv.org/abs/astro-ph/0603494}{{\ttfamily
  astro-ph/0603494}}].

\bibitem{Planck:2018vyg}
{\scshape Planck} collaboration, \emph{{Planck 2018 results. VI. Cosmological
  parameters}},
  \href{https://doi.org/10.1051/0004-6361/201833910}{\emph{Astron. Astrophys.}
  {\bfseries 641} (2020) A6}
  [\href{https://arxiv.org/abs/1807.06209}{{\ttfamily 1807.06209}}].

\bibitem{DESI:2024mwx}
{\scshape DESI} collaboration, \emph{{DESI 2024 VI: Cosmological Constraints
  from the Measurements of Baryon Acoustic Oscillations}},
  \href{https://arxiv.org/abs/2404.03002}{{\ttfamily 2404.03002}}.

\bibitem{Esteban:2024eli}
I.~Esteban, M.C.~Gonzalez-Garcia, M.~Maltoni, I.~Martinez-Soler,
  J.a.P.~Pinheiro and T.~Schwetz, \emph{{NuFit-6.0: Updated global analysis of
  three-flavor neutrino oscillations}},
  \href{https://arxiv.org/abs/2410.05380}{{\ttfamily 2410.05380}}.

\bibitem{Craig:2024tky}
N.~Craig, D.~Green, J.~Meyers and S.~Rajendran, \emph{{No \ensuremath{\nu}s is
  Good News}}, \href{https://doi.org/10.1007/JHEP09(2024)097}{\emph{JHEP}
  {\bfseries 09} (2024) 097}
  [\href{https://arxiv.org/abs/2405.00836}{{\ttfamily 2405.00836}}].

\bibitem{Green:2024xbb}
D.~Green and J.~Meyers, \emph{{The Cosmological Preference for Negative
  Neutrino Mass}},  \href{https://arxiv.org/abs/2407.07878}{{\ttfamily
  2407.07878}}.

\bibitem{Elbers:2024sha}
W.~Elbers, C.S.~Frenk, A.~Jenkins, B.~Li and S.~Pascoli, \emph{{Negative
  neutrino masses as a mirage of dark energy}},
  \href{https://arxiv.org/abs/2407.10965}{{\ttfamily 2407.10965}}.

\bibitem{Naredo-Tuero:2024sgf}
D.~Naredo-Tuero, M.~Escudero, E.~Fern\'andez-Mart\'\i{}nez, X.~Marcano and
  V.~Poulin, \emph{{Living at the Edge: A Critical Look at the Cosmological
  Neutrino Mass Bound}},  \href{https://arxiv.org/abs/2407.13831}{{\ttfamily
  2407.13831}}.

\bibitem{Loverde:2024nfi}
M.~Loverde and Z.J.~Weiner, \emph{{Massive neutrinos and cosmic composition}},
  \href{https://arxiv.org/abs/2410.00090}{{\ttfamily 2410.00090}}.

\bibitem{RoyChoudhury:2024wri}
S.~Roy~Choudhury and T.~Okumura, \emph{{Updated cosmological constraints in
  extended parameter space with Planck PR4, DESI BAO, and SN: dynamical dark
  energy, neutrino masses, lensing anomaly, and the Hubble tension}},
  \href{https://arxiv.org/abs/2409.13022}{{\ttfamily 2409.13022}}.

\bibitem{Gariazzo:2023joe}
S.~Gariazzo, O.~Mena and T.~Schwetz, \emph{{Quantifying the tension between
  cosmological and terrestrial constraints on neutrino masses}},
  \href{https://doi.org/10.1016/j.dark.2023.101226}{\emph{Phys. Dark Univ.}
  {\bfseries 40} (2023) 101226}
  [\href{https://arxiv.org/abs/2302.14159}{{\ttfamily 2302.14159}}].

\bibitem{Katrin:2024tvg}
{\scshape Katrin} collaboration, \emph{{Direct neutrino-mass measurement based
  on 259 days of KATRIN data}},
  \href{https://arxiv.org/abs/2406.13516}{{\ttfamily 2406.13516}}.

\bibitem{KamLAND-Zen:2024eml}
{\scshape KamLAND-Zen} collaboration, \emph{{Search for Majorana Neutrinos with
  the Complete KamLAND-Zen Dataset}},
  \href{https://arxiv.org/abs/2406.11438}{{\ttfamily 2406.11438}}.

\bibitem{Escudero:2019gfk}
M.~Escudero and M.~Fairbairn, \emph{{Cosmological Constraints on Invisible
  Neutrino Decays Revisited}},
  \href{https://doi.org/10.1103/PhysRevD.100.103531}{\emph{Phys. Rev. D}
  {\bfseries 100} (2019) 103531}
  [\href{https://arxiv.org/abs/1907.05425}{{\ttfamily 1907.05425}}].

\bibitem{Chacko:2019nej}
Z.~Chacko, A.~Dev, P.~Du, V.~Poulin and Y.~Tsai, \emph{{Cosmological Limits on
  the Neutrino Mass and Lifetime}},
  \href{https://doi.org/10.1007/JHEP04(2020)020}{\emph{JHEP} {\bfseries 04}
  (2020) 020} [\href{https://arxiv.org/abs/1909.05275}{{\ttfamily
  1909.05275}}].

\bibitem{Escudero:2020ped}
M.~Escudero, J.~Lopez-Pavon, N.~Rius and S.~Sandner, \emph{{Relaxing
  Cosmological Neutrino Mass Bounds with Unstable Neutrinos}},
  \href{https://doi.org/10.1007/JHEP12(2020)119}{\emph{JHEP} {\bfseries 12}
  (2020) 119} [\href{https://arxiv.org/abs/2007.04994}{{\ttfamily
  2007.04994}}].

\bibitem{Chacko:2020hmh}
Z.~Chacko, A.~Dev, P.~Du, V.~Poulin and Y.~Tsai, \emph{{Determining the
  Neutrino Lifetime from Cosmology}},
  \href{https://doi.org/10.1103/PhysRevD.103.043519}{\emph{Phys. Rev. D}
  {\bfseries 103} (2021) 043519}
  [\href{https://arxiv.org/abs/2002.08401}{{\ttfamily 2002.08401}}].

\bibitem{Barenboim:2020vrr}
G.~Barenboim, J.Z.~Chen, S.~Hannestad, I.M.~Oldengott, T.~Tram and Y.Y.Y.~Wong,
  \emph{{Invisible neutrino decay in precision cosmology}},
  \href{https://doi.org/10.1088/1475-7516/2021/03/087}{\emph{JCAP} {\bfseries
  03} (2021) 087} [\href{https://arxiv.org/abs/2011.01502}{{\ttfamily
  2011.01502}}].

\bibitem{FrancoAbellan:2021hdb}
G.~Franco~Abell\'an, Z.~Chacko, A.~Dev, P.~Du, V.~Poulin and Y.~Tsai,
  \emph{{Improved cosmological constraints on the neutrino mass and lifetime}},
  \href{https://doi.org/10.1007/JHEP08(2022)076}{\emph{JHEP} {\bfseries 08}
  (2022) 076} [\href{https://arxiv.org/abs/2112.13862}{{\ttfamily
  2112.13862}}].

\bibitem{Chen:2022idm}
J.Z.~Chen, I.M.~Oldengott, G.~Pierobon and Y.Y.Y.~Wong, \emph{{Weaker yet
  again: mass spectrum-consistent cosmological constraints on the neutrino
  lifetime}}, \href{https://doi.org/10.1140/epjc/s10052-022-10518-3}{\emph{Eur.
  Phys. J. C} {\bfseries 82} (2022) 640}
  [\href{https://arxiv.org/abs/2203.09075}{{\ttfamily 2203.09075}}].

\bibitem{Dvali:2016uhn}
G.~Dvali and L.~Funcke, \emph{{Small neutrino masses from gravitational
  \ensuremath{\theta}-term}},
  \href{https://doi.org/10.1103/PhysRevD.93.113002}{\emph{Phys. Rev. D}
  {\bfseries 93} (2016) 113002}
  [\href{https://arxiv.org/abs/1602.03191}{{\ttfamily 1602.03191}}].

\bibitem{Lorenz:2018fzb}
C.S.~Lorenz, L.~Funcke, E.~Calabrese and S.~Hannestad, \emph{{Time-varying
  neutrino mass from a supercooled phase transition: current cosmological
  constraints and impact on the $\Omega_m$-$\sigma_8$ plane}},
  \href{https://doi.org/10.1103/PhysRevD.99.023501}{\emph{Phys. Rev. D}
  {\bfseries 99} (2019) 023501}
  [\href{https://arxiv.org/abs/1811.01991}{{\ttfamily 1811.01991}}].

\bibitem{Dvali:2021uvk}
G.~Dvali, L.~Funcke and T.~Vachaspati, \emph{{Time- and Space-Varying Neutrino
  Mass Matrix from Soft Topological Defects}},
  \href{https://doi.org/10.1103/PhysRevLett.130.091601}{\emph{Phys. Rev. Lett.}
  {\bfseries 130} (2023) 091601}
  [\href{https://arxiv.org/abs/2112.02107}{{\ttfamily 2112.02107}}].

\bibitem{Lorenz:2021alz}
C.S.~Lorenz, L.~Funcke, M.~L\"offler and E.~Calabrese, \emph{{Reconstruction of
  the neutrino mass as a function of redshift}},
  \href{https://doi.org/10.1103/PhysRevD.104.123518}{\emph{Phys. Rev. D}
  {\bfseries 104} (2021) 123518}
  [\href{https://arxiv.org/abs/2102.13618}{{\ttfamily 2102.13618}}].

\bibitem{Esteban:2021ozz}
I.~Esteban and J.~Salvado, \emph{{Long Range Interactions in Cosmology:
  Implications for Neutrinos}},
  \href{https://doi.org/10.1088/1475-7516/2021/05/036}{\emph{JCAP} {\bfseries
  05} (2021) 036} [\href{https://arxiv.org/abs/2101.05804}{{\ttfamily
  2101.05804}}].

\bibitem{Sen:2023uga}
M.~Sen and A.Y.~Smirnov, \emph{{Refractive neutrino masses, ultralight dark
  matter and cosmology}},
  \href{https://doi.org/10.1088/1475-7516/2024/01/040}{\emph{JCAP} {\bfseries
  01} (2024) 040} [\href{https://arxiv.org/abs/2306.15718}{{\ttfamily
  2306.15718}}].

\bibitem{Escudero:2022gez}
M.~Escudero, T.~Schwetz and J.~Terol-Calvo, \emph{{A seesaw model for large
  neutrino masses in concordance with cosmology}},
  \href{https://doi.org/10.1007/JHEP02(2023)142}{\emph{JHEP} {\bfseries 02}
  (2023) 142} [\href{https://arxiv.org/abs/2211.01729}{{\ttfamily
  2211.01729}}].

\bibitem{Farzan:2015pca}
Y.~Farzan and S.~Hannestad, \emph{{Neutrinos secretly converting to lighter
  particles to please both KATRIN and the cosmos}},
  \href{https://doi.org/10.1088/1475-7516/2016/02/058}{\emph{JCAP} {\bfseries
  02} (2016) 058} [\href{https://arxiv.org/abs/1510.02201}{{\ttfamily
  1510.02201}}].

\bibitem{Allali:2024anb}
I.J.~Allali, D.~Aloni and N.~Sch\"oneberg, \emph{{Cosmological probes of Dark
  Radiation from Neutrino Mixing}},
  \href{https://doi.org/10.1088/1475-7516/2024/09/019}{\emph{JCAP} {\bfseries
  09} (2024) 019} [\href{https://arxiv.org/abs/2404.16822}{{\ttfamily
  2404.16822}}].

\bibitem{GAMBITCosmologyWorkgroup:2020htv}
{\scshape GAMBIT Cosmology Workgroup} collaboration, \emph{{CosmoBit: A GAMBIT
  module for computing cosmological observables and likelihoods}},
  \href{https://doi.org/10.1088/1475-7516/2021/02/022}{\emph{JCAP} {\bfseries
  02} (2021) 022} [\href{https://arxiv.org/abs/2009.03286}{{\ttfamily
  2009.03286}}].

\bibitem{Cuoco:2005qr}
A.~Cuoco, J.~Lesgourgues, G.~Mangano and S.~Pastor, \emph{{Do observations
  prove that cosmological neutrinos are thermally distributed?}},
  \href{https://doi.org/10.1103/PhysRevD.71.123501}{\emph{Phys. Rev. D}
  {\bfseries 71} (2005) 123501}
  [\href{https://arxiv.org/abs/astro-ph/0502465}{{\ttfamily
  astro-ph/0502465}}].

\bibitem{Alvey:2021sji}
J.~Alvey, M.~Escudero and N.~Sabti, \emph{{What can CMB observations tell us
  about the neutrino distribution function?}},
  \href{https://doi.org/10.1088/1475-7516/2022/02/037}{\emph{JCAP} {\bfseries
  02} (2022) 037} [\href{https://arxiv.org/abs/2111.12726}{{\ttfamily
  2111.12726}}].

\bibitem{Oldengott:2019lke}
I.M.~Oldengott, G.~Barenboim, S.~Kahlen, J.~Salvado and D.J.~Schwarz,
  \emph{{How to relax the cosmological neutrino mass bound}},
  \href{https://doi.org/10.1088/1475-7516/2019/04/049}{\emph{JCAP} {\bfseries
  04} (2019) 049} [\href{https://arxiv.org/abs/1901.04352}{{\ttfamily
  1901.04352}}].

\bibitem{Alvey:2021xmq}
J.~Alvey, M.~Escudero, N.~Sabti and T.~Schwetz, \emph{{Cosmic neutrino
  background detection in large-neutrino-mass cosmologies}},
  \href{https://doi.org/10.1103/PhysRevD.105.063501}{\emph{Phys. Rev. D}
  {\bfseries 105} (2022) 063501}
  [\href{https://arxiv.org/abs/2111.14870}{{\ttfamily 2111.14870}}].

\bibitem{Drewes:2016upu}
M.~Drewes et~al., \emph{{A White Paper on keV Sterile Neutrino Dark Matter}},
  \href{https://doi.org/10.1088/1475-7516/2017/01/025}{\emph{JCAP} {\bfseries
  01} (2017) 025} [\href{https://arxiv.org/abs/1602.04816}{{\ttfamily
  1602.04816}}].

\bibitem{Abazajian:2017tcc}
K.N.~Abazajian, \emph{{Sterile neutrinos in cosmology}},
  \href{https://doi.org/10.1016/j.physrep.2017.10.003}{\emph{Phys. Rept.}
  {\bfseries 711-712} (2017) 1}
  [\href{https://arxiv.org/abs/1705.01837}{{\ttfamily 1705.01837}}].

\bibitem{Boyarsky:2018tvu}
A.~Boyarsky, M.~Drewes, T.~Lasserre, S.~Mertens and O.~Ruchayskiy,
  \emph{{Sterile neutrino Dark Matter}},
  \href{https://doi.org/10.1016/j.ppnp.2018.07.004}{\emph{Prog. Part. Nucl.
  Phys.} {\bfseries 104} (2019) 1}
  [\href{https://arxiv.org/abs/1807.07938}{{\ttfamily 1807.07938}}].

\bibitem{DeGouvea:2019wpf}
A.~De~Gouv\^ea, M.~Sen, W.~Tangarife and Y.~Zhang, \emph{{Dodelson-Widrow
  Mechanism in the Presence of Self-Interacting Neutrinos}},
  \href{https://doi.org/10.1103/PhysRevLett.124.081802}{\emph{Phys. Rev. Lett.}
  {\bfseries 124} (2020) 081802}
  [\href{https://arxiv.org/abs/1910.04901}{{\ttfamily 1910.04901}}].

\bibitem{Kelly:2020pcy}
K.J.~Kelly, M.~Sen, W.~Tangarife and Y.~Zhang, \emph{{Origin of sterile
  neutrino dark matter via secret neutrino interactions with vector bosons}},
  \href{https://doi.org/10.1103/PhysRevD.101.115031}{\emph{Phys. Rev. D}
  {\bfseries 101} (2020) 115031}
  [\href{https://arxiv.org/abs/2005.03681}{{\ttfamily 2005.03681}}].

\bibitem{Benso:2021hhh}
C.~Benso, W.~Rodejohann, M.~Sen and A.U.~Ramachandran, \emph{{Sterile neutrino
  dark matter production in presence of nonstandard neutrino self-interactions:
  An EFT approach}},
  \href{https://doi.org/10.1103/PhysRevD.105.055016}{\emph{Phys. Rev. D}
  {\bfseries 105} (2022) 055016}
  [\href{https://arxiv.org/abs/2112.00758}{{\ttfamily 2112.00758}}].

\bibitem{Johns:2019cwc}
L.~Johns and G.M.~Fuller, \emph{{Self-interacting sterile neutrino dark matter:
  the heavy-mediator case}},
  \href{https://doi.org/10.1103/PhysRevD.100.023533}{\emph{Phys. Rev. D}
  {\bfseries 100} (2019) 023533}
  [\href{https://arxiv.org/abs/1903.08296}{{\ttfamily 1903.08296}}].

\bibitem{Bringmann:2022aim}
T.~Bringmann, P.F.~Depta, M.~Hufnagel, J.~Kersten, J.T.~Ruderman and
  K.~Schmidt-Hoberg, \emph{{Minimal sterile neutrino dark matter}},
  \href{https://doi.org/10.1103/PhysRevD.107.L071702}{\emph{Phys. Rev. D}
  {\bfseries 107} (2023) L071702}
  [\href{https://arxiv.org/abs/2206.10630}{{\ttfamily 2206.10630}}].

\bibitem{Astros:2023xhe}
M.D.~Astros and S.~Vogl, \emph{{Boosting the production of sterile neutrino
  dark matter with self-interactions}},
  \href{https://doi.org/10.1007/JHEP03(2024)032}{\emph{JHEP} {\bfseries 03}
  (2024) 032} [\href{https://arxiv.org/abs/2307.15565}{{\ttfamily
  2307.15565}}].

\bibitem{Fuller:2024noz}
G.M.~Fuller, L.~Gr\'af, A.V.~Patwardhan and J.~Spisak, \emph{{Dark population
  transfer mechanism for sterile neutrino dark matter}},
  \href{https://arxiv.org/abs/2402.13878}{{\ttfamily 2402.13878}}.

\bibitem{Datta:2021elq}
A.~Datta, R.~Roshan and A.~Sil, \emph{{Imprint of the Seesaw Mechanism on
  Feebly Interacting Dark Matter and the Baryon Asymmetry}},
  \href{https://doi.org/10.1103/PhysRevLett.127.231801}{\emph{Phys. Rev. Lett.}
  {\bfseries 127} (2021) 231801}
  [\href{https://arxiv.org/abs/2104.02030}{{\ttfamily 2104.02030}}].

\bibitem{Goertz:2024gzw}
F.~Goertz, M.~Hager, G.~Laverda and J.~Rubio, \emph{{Phasing out of Darkness:
  From Sterile Neutrino Dark Matter to Neutrino Masses via Time-Dependent
  Mixing}},  \href{https://arxiv.org/abs/2407.04778}{{\ttfamily 2407.04778}}.

\bibitem{Berlin:2017ftj}
A.~Berlin and N.~Blinov, \emph{{Thermal Dark Matter Below an MeV}},
  \href{https://doi.org/10.1103/PhysRevLett.120.021801}{\emph{Phys. Rev. Lett.}
  {\bfseries 120} (2018) 021801}
  [\href{https://arxiv.org/abs/1706.07046}{{\ttfamily 1706.07046}}].

\bibitem{Berlin:2018ztp}
A.~Berlin and N.~Blinov, \emph{{Thermal neutrino portal to sub-MeV dark
  matter}}, \href{https://doi.org/10.1103/PhysRevD.99.095030}{\emph{Phys. Rev.
  D} {\bfseries 99} (2019) 095030}
  [\href{https://arxiv.org/abs/1807.04282}{{\ttfamily 1807.04282}}].

\bibitem{Barry:2011wb}
J.~Barry, W.~Rodejohann and H.~Zhang, \emph{{Light Sterile Neutrinos: Models
  and Phenomenology}},
  \href{https://doi.org/10.1007/JHEP07(2011)091}{\emph{JHEP} {\bfseries 07}
  (2011) 091} [\href{https://arxiv.org/abs/1105.3911}{{\ttfamily 1105.3911}}].

\bibitem{EscuderoAbenza:2020cmq}
M.~Escudero~Abenza, \emph{{Precision early universe thermodynamics made simple:
  $N_{\rm eff}$ and neutrino decoupling in the Standard Model and beyond}},
  \href{https://doi.org/10.1088/1475-7516/2020/05/048}{\emph{JCAP} {\bfseries
  05} (2020) 048} [\href{https://arxiv.org/abs/2001.04466}{{\ttfamily
  2001.04466}}].

\bibitem{Akita:2020szl}
K.~Akita and M.~Yamaguchi, \emph{{A precision calculation of relic neutrino
  decoupling}},
  \href{https://doi.org/10.1088/1475-7516/2020/08/012}{\emph{JCAP} {\bfseries
  08} (2020) 012} [\href{https://arxiv.org/abs/2005.07047}{{\ttfamily
  2005.07047}}].

\bibitem{ParticleDataGroup:2024cfk}
{\scshape Particle Data Group} collaboration, \emph{{Review of particle
  physics}}, \href{https://doi.org/10.1103/PhysRevD.110.030001}{\emph{Phys.
  Rev. D} {\bfseries 110} (2024) 030001}.

\bibitem{Belanger:2011ww}
G.~Belanger and J.-C.~Park, \emph{{Assisted freeze-out}},
  \href{https://doi.org/10.1088/1475-7516/2012/03/038}{\emph{JCAP} {\bfseries
  03} (2012) 038} [\href{https://arxiv.org/abs/1112.4491}{{\ttfamily
  1112.4491}}].

\bibitem{Liu:2011aa}
Z.-P.~Liu, Y.-L.~Wu and Y.-F.~Zhou, \emph{{Enhancement of dark matter relic
  density from the late time dark matter conversions}},
  \href{https://doi.org/10.1140/epjc/s10052-011-1749-4}{\emph{Eur. Phys. J. C}
  {\bfseries 71} (2011) 1749}
  [\href{https://arxiv.org/abs/1101.4148}{{\ttfamily 1101.4148}}].

\bibitem{BasiBeneito:2022qxd}
A.~Bas~i Beneito, J.~Herrero-Garc\'\i{}a and D.~Vatsyayan,
  \emph{{Multi-component dark sectors: symmetries, asymmetries and
  conversions}}, \href{https://doi.org/10.1007/JHEP10(2022)075}{\emph{JHEP}
  {\bfseries 10} (2022) 075}
  [\href{https://arxiv.org/abs/2207.02874}{{\ttfamily 2207.02874}}].

\bibitem{Ko:2014bka}
P.~Ko and Y.~Tang, \emph{{\ensuremath{\nu}\ensuremath{\Lambda}MDM: A model for
  sterile neutrino and dark matter reconciles cosmological and neutrino
  oscillation data after BICEP2}},
  \href{https://doi.org/10.1016/j.physletb.2014.10.035}{\emph{Phys. Lett. B}
  {\bfseries 739} (2014) 62} [\href{https://arxiv.org/abs/1404.0236}{{\ttfamily
  1404.0236}}].

\bibitem{Laha:2013xua}
R.~Laha, B.~Dasgupta and J.F.~Beacom, \emph{{Constraints on New Neutrino
  Interactions via Light Abelian Vector Bosons}},
  \href{https://doi.org/10.1103/PhysRevD.89.093025}{\emph{Phys. Rev. D}
  {\bfseries 89} (2014) 093025}
  [\href{https://arxiv.org/abs/1304.3460}{{\ttfamily 1304.3460}}].

\bibitem{Bakhti:2017jhm}
P.~Bakhti and Y.~Farzan, \emph{{Constraining secret gauge interactions of
  neutrinos by meson decays}},
  \href{https://doi.org/10.1103/PhysRevD.95.095008}{\emph{Phys. Rev. D}
  {\bfseries 95} (2017) 095008}
  [\href{https://arxiv.org/abs/1702.04187}{{\ttfamily 1702.04187}}].

\bibitem{Arcadi:2018xdd}
G.~Arcadi, J.~Heeck, F.~Heizmann, S.~Mertens, F.S.~Queiroz, W.~Rodejohann
  et~al., \emph{{Tritium beta decay with additional emission of new light
  bosons}}, \href{https://doi.org/10.1007/JHEP01(2019)206}{\emph{JHEP}
  {\bfseries 01} (2019) 206}
  [\href{https://arxiv.org/abs/1811.03530}{{\ttfamily 1811.03530}}].

\bibitem{Kelly:2019wow}
K.J.~Kelly and Y.~Zhang, \emph{{Mononeutrino at DUNE: New Signals from
  Neutrinophilic Thermal Dark Matter}},
  \href{https://doi.org/10.1103/PhysRevD.99.055034}{\emph{Phys. Rev. D}
  {\bfseries 99} (2019) 055034}
  [\href{https://arxiv.org/abs/1901.01259}{{\ttfamily 1901.01259}}].

\bibitem{Brdar:2020nbj}
V.~Brdar, M.~Lindner, S.~Vogl and X.-J.~Xu, \emph{{Revisiting neutrino
  self-interaction constraints from $Z$ and $\tau$ decays}},
  \href{https://doi.org/10.1103/PhysRevD.101.115001}{\emph{Phys. Rev. D}
  {\bfseries 101} (2020) 115001}
  [\href{https://arxiv.org/abs/2003.05339}{{\ttfamily 2003.05339}}].

\bibitem{Dev:2024twk}
P.S.B.~Dev, D.~Kim, D.~Sathyan, K.~Sinha and Y.~Zhang, \emph{{New Laboratory
  Constraints on Neutrinophilic Mediators}},
  \href{https://arxiv.org/abs/2407.12738}{{\ttfamily 2407.12738}}.

\bibitem{Feng:2008mu}
J.L.~Feng, H.~Tu and H.-B.~Yu, \emph{{Thermal Relics in Hidden Sectors}},
  \href{https://doi.org/10.1088/1475-7516/2008/10/043}{\emph{JCAP} {\bfseries
  10} (2008) 043} [\href{https://arxiv.org/abs/0808.2318}{{\ttfamily
  0808.2318}}].

\bibitem{Berlin:2016gtr}
A.~Berlin, D.~Hooper and G.~Krnjaic, \emph{{Thermal Dark Matter From A Highly
  Decoupled Sector}},
  \href{https://doi.org/10.1103/PhysRevD.94.095019}{\emph{Phys. Rev. D}
  {\bfseries 94} (2016) 095019}
  [\href{https://arxiv.org/abs/1609.02555}{{\ttfamily 1609.02555}}].

\bibitem{Berlin:2014tja}
A.~Berlin, D.~Hooper and S.D.~McDermott, \emph{{Simplified Dark Matter Models
  for the Galactic Center Gamma-Ray Excess}},
  \href{https://doi.org/10.1103/PhysRevD.89.115022}{\emph{Phys. Rev. D}
  {\bfseries 89} (2014) 115022}
  [\href{https://arxiv.org/abs/1404.0022}{{\ttfamily 1404.0022}}].

\bibitem{Arcadi:2024ukq}
G.~Arcadi, D.~Cabo-Almeida, M.~Dutra, P.~Ghosh, M.~Lindner, Y.~Mambrini et~al.,
  \emph{{The Waning of the WIMP: Endgame?}},
  \href{https://arxiv.org/abs/2403.15860}{{\ttfamily 2403.15860}}.

\bibitem{Kolb:1990vq}
E.W.~Kolb, \emph{{The Early Universe}}, vol.~69, Taylor and Francis (5, 2019),
  \href{https://doi.org/10.1201/9780429492860}{10.1201/9780429492860}.

\bibitem{Audren:2014bca}
B.~Audren, J.~Lesgourgues, G.~Mangano, P.D.~Serpico and T.~Tram,
  \emph{{Strongest model-independent bound on the lifetime of Dark Matter}},
  \href{https://doi.org/10.1088/1475-7516/2014/12/028}{\emph{JCAP} {\bfseries
  12} (2014) 028} [\href{https://arxiv.org/abs/1407.2418}{{\ttfamily
  1407.2418}}].

\bibitem{Poulin:2016nat}
V.~Poulin, P.D.~Serpico and J.~Lesgourgues, \emph{{A fresh look at linear
  cosmological constraints on a decaying dark matter component}},
  \href{https://doi.org/10.1088/1475-7516/2016/08/036}{\emph{JCAP} {\bfseries
  08} (2016) 036} [\href{https://arxiv.org/abs/1606.02073}{{\ttfamily
  1606.02073}}].

\bibitem{Abazajian:2021zui}
K.N.~Abazajian, \emph{{Neutrinos in Astrophysics and Cosmology: Theoretical
  Advanced Study Institute (TASI) 2020 Lectures}},
  \href{https://doi.org/10.22323/1.388.0001}{\emph{PoS} {\bfseries TASI2020}
  (2021) 001} [\href{https://arxiv.org/abs/2102.10183}{{\ttfamily
  2102.10183}}].

\bibitem{Bulbul:2014sua}
E.~Bulbul, M.~Markevitch, A.~Foster, R.K.~Smith, M.~Loewenstein and
  S.W.~Randall, \emph{{Detection of An Unidentified Emission Line in the
  Stacked X-ray spectrum of Galaxy Clusters}},
  \href{https://doi.org/10.1088/0004-637X/789/1/13}{\emph{Astrophys. J.}
  {\bfseries 789} (2014) 13} [\href{https://arxiv.org/abs/1402.2301}{{\ttfamily
  1402.2301}}].

\bibitem{Boyarsky:2014jta}
A.~Boyarsky, O.~Ruchayskiy, D.~Iakubovskyi and J.~Franse, \emph{{Unidentified
  Line in X-Ray Spectra of the Andromeda Galaxy and Perseus Galaxy Cluster}},
  \href{https://doi.org/10.1103/PhysRevLett.113.251301}{\emph{Phys. Rev. Lett.}
  {\bfseries 113} (2014) 251301}
  [\href{https://arxiv.org/abs/1402.4119}{{\ttfamily 1402.4119}}].

\bibitem{Boyarsky:2014ska}
A.~Boyarsky, J.~Franse, D.~Iakubovskyi and O.~Ruchayskiy, \emph{{Checking the
  Dark Matter Origin of a 3.53 keV Line with the Milky Way Center}},
  \href{https://doi.org/10.1103/PhysRevLett.115.161301}{\emph{Phys. Rev. Lett.}
  {\bfseries 115} (2015) 161301}
  [\href{https://arxiv.org/abs/1408.2503}{{\ttfamily 1408.2503}}].

\bibitem{Dessert:2023fen}
C.~Dessert, J.W.~Foster, Y.~Park and B.R.~Safdi, \emph{{Was There a 3.5 keV
  Line?}}, \href{https://doi.org/10.3847/1538-4357/ad2612}{\emph{Astrophys. J.}
  {\bfseries 964} (2024) 185}
  [\href{https://arxiv.org/abs/2309.03254}{{\ttfamily 2309.03254}}].

\bibitem{Pal:1981rm}
P.B.~Pal and L.~Wolfenstein, \emph{{Radiative Decays of Massive Neutrinos}},
  \href{https://doi.org/10.1103/PhysRevD.25.766}{\emph{Phys. Rev. D} {\bfseries
  25} (1982) 766}.

\bibitem{An:2023mkf}
R.~An, V.~Gluscevic, E.O.~Nadler and Y.~Zhang, \emph{{Can Neutrino
  Self-interactions Save Sterile Neutrino Dark Matter?}},
  \href{https://doi.org/10.3847/2041-8213/acf049}{\emph{Astrophys. J. Lett.}
  {\bfseries 954} (2023) L18}
  [\href{https://arxiv.org/abs/2301.08299}{{\ttfamily 2301.08299}}].

\bibitem{Zelko:2022tgf}
I.A.~Zelko, T.~Treu, K.N.~Abazajian, D.~Gilman, A.J.~Benson, S.~Birrer et~al.,
  \emph{{Constraints on Sterile Neutrino Models from Strong Gravitational
  Lensing, Milky~Way Satellites, and the Lyman-\ensuremath{\alpha} Forest}},
  \href{https://doi.org/10.1103/PhysRevLett.129.191301}{\emph{Phys. Rev. Lett.}
  {\bfseries 129} (2022) 191301}
  [\href{https://arxiv.org/abs/2205.09777}{{\ttfamily 2205.09777}}].

\bibitem{Enzi:2020ieg}
W.~Enzi et~al., \emph{{Joint constraints on thermal relic dark matter from
  strong gravitational lensing, the Ly\,\ensuremath{\alpha} forest, and Milky
  Way satellites}}, \href{https://doi.org/10.1093/mnras/stab1960}{\emph{Mon.
  Not. Roy. Astron. Soc.} {\bfseries 506} (2021) 5848}
  [\href{https://arxiv.org/abs/2010.13802}{{\ttfamily 2010.13802}}].

\bibitem{Bringmann:2016ilk}
T.~Bringmann, H.T.~Ihle, J.~Kersten and P.~Walia, \emph{{Suppressing structure
  formation at dwarf galaxy scales and below: late kinetic decoupling as a
  compelling alternative to warm dark matter}},
  \href{https://doi.org/10.1103/PhysRevD.94.103529}{\emph{Phys. Rev. D}
  {\bfseries 94} (2016) 103529}
  [\href{https://arxiv.org/abs/1603.04884}{{\ttfamily 1603.04884}}].

\bibitem{Cyr-Racine:2013fsa}
F.-Y.~Cyr-Racine, R.~de~Putter, A.~Raccanelli and K.~Sigurdson,
  \emph{{Constraints on Large-Scale Dark Acoustic Oscillations from
  Cosmology}}, \href{https://doi.org/10.1103/PhysRevD.89.063517}{\emph{Phys.
  Rev. D} {\bfseries 89} (2014) 063517}
  [\href{https://arxiv.org/abs/1310.3278}{{\ttfamily 1310.3278}}].

\bibitem{Taule:2022jrz}
P.~Taule, M.~Escudero and M.~Garny, \emph{{Global view of neutrino interactions
  in cosmology: The free streaming window as seen by Planck}},
  \href{https://doi.org/10.1103/PhysRevD.106.063539}{\emph{Phys. Rev. D}
  {\bfseries 106} (2022) 063539}
  [\href{https://arxiv.org/abs/2207.04062}{{\ttfamily 2207.04062}}].

\bibitem{cite-key}
M.~Escudero, T.~Schwetz and J.~Terol-Calvo, \emph{Addendum to: A seesaw model
  for large neutrino masses in concordance with cosmology},
  \href{https://doi.org/10.1007/JHEP06(2024)119}{\emph{Journal of High Energy
  Physics} {\bfseries 2024} (2024) 119}.

\bibitem{Abazajian:2019eic}
K.~Abazajian et~al., \emph{{CMB-S4 Science Case, Reference Design, and Project
  Plan}},  \href{https://arxiv.org/abs/1907.04473}{{\ttfamily 1907.04473}}.

\bibitem{SimonsObservatory:2018koc}
{\scshape Simons Observatory} collaboration, \emph{{The Simons Observatory:
  Science goals and forecasts}},
  \href{https://doi.org/10.1088/1475-7516/2019/02/056}{\emph{JCAP} {\bfseries
  02} (2019) 056} [\href{https://arxiv.org/abs/1808.07445}{{\ttfamily
  1808.07445}}].

\bibitem{Upadhye:2024ypg}
A.~Upadhye, M.R.~Mosbech, G.~Pierobon and Y.Y.Y.~Wong, \emph{{Everything hot
  everywhere all at once: Neutrinos and hot dark matter as a single effective
  species}},  \href{https://arxiv.org/abs/2410.05815}{{\ttfamily 2410.05815}}.

\bibitem{Allali:2024cji}
I.J.~Allali, A.~Notari and F.~Rompineve, \emph{{Dark Radiation with Baryon
  Acoustic Oscillations from DESI 2024 and the $H_0$ tension}},
  \href{https://arxiv.org/abs/2404.15220}{{\ttfamily 2404.15220}}.

\bibitem{Grimus:2000vj}
W.~Grimus and L.~Lavoura, \emph{{The Seesaw mechanism at arbitrary order:
  Disentangling the small scale from the large scale}},
  \href{https://doi.org/10.1088/1126-6708/2000/11/042}{\emph{JHEP} {\bfseries
  11} (2000) 042} [\href{https://arxiv.org/abs/hep-ph/0008179}{{\ttfamily
  hep-ph/0008179}}].

\end{thebibliography}\endgroup

\end{document}